\documentclass[sigconf, screen]{acmart}

\AtBeginDocument{%
  }

\copyrightyear{2026}
\acmYear{2026}
\setcopyright{cc}
\setcctype{by-nc-nd}
\acmConference[CHI '26]{Proceedings of the 2026 CHI Conference on Human Factors in Computing Systems}{April 13--17, 2026}{Barcelona, Spain}
\acmBooktitle{Proceedings of the 2026 CHI Conference on Human Factors in Computing Systems (CHI '26), April 13--17, 2026, Barcelona, Spain}
\acmDOI{10.1145/3772318.3791195}
\acmISBN{979-8-4007-2278-3/2026/04}

\usepackage{enumitem} 
\usepackage{multirow} 
\usepackage{rotating} 
\usepackage{comment} 
\usepackage{float}
\usepackage{microtype}
\usepackage{xspace}

\usepackage{colortbl}

\usepackage{stfloats}

\newcommand{\m}{\textit{M=}}

\newcommand{\sd}{\textit{SD=}}

\newcommand{\chall}[1]{\hyperref[challenge:#1]{\textbf{#1}}}

\newcommand\itema{\item[\textbf{o1}]}
\newcommand\itemb{\item[\textbf{o2}]}
\newcommand\itemc{\item[\textbf{o3}]}
\newcommand\itemd{\item[\textbf{o4}]}

\newcommand\itemrq{\item[\textbf{RQ}]}

\newcommand{\tool}{\textsc{MIRAGE}\xspace}

\usepackage{pifont}

\begin{document}

\title{\tool: Enabling Real-Time Automotive Mediated Reality}

\author{Pascal Jansen}
\authornote{Authors contributed equally to this research.}
\email{pascal.jansen@uni-ulm.de}
\orcid{0000-0002-9335-5462}
\affiliation{%
  \institution{Institute of Media Informatics, Ulm University}
  \city{Ulm}
  \country{Germany}
}
\affiliation{%
  \institution{UCL Interaction Centre}
  \city{London}
  \country{UK}
}

\author{Julian Britten}
\authornotemark[1]
\email{julian.britten@uni-ulm.de}
\orcid{0000-0002-2646-2727}
\affiliation{%
  \institution{Institute of Media Informatics, Ulm University}
  \city{Ulm}
  \country{Germany}
}

\author{Mark Colley}
\authornotemark[1]
\email{m.colley@ucl.ac.uk}
\orcid{0000-0001-5207-5029}
\affiliation{%
  \institution{UCL Interaction Centre}
  \city{London}
  \country{UK}
}

\author{Markus Sasalovici}
\email{markus.sasalovici@mercedes-benz.com}
\orcid{0000-0001-9883-2398}
\affiliation{%
    \institution{Mercedes-Benz Tech Innovation GmbH}
  \institution{Institute of Media Informatics, Ulm University}
  \city{Ulm}
  \country{Germany}
}

\author{Enrico Rukzio}
\email{enrico.rukzio@uni-ulm.de}
\orcid{0000-0002-4213-2226}
\affiliation{%
  \institution{Institute of Media Informatics, Ulm University}
  \city{Ulm}
  \country{Germany}
}

\renewcommand{\shortauthors}{Jansen, Britten, Colley et al.}

\begin{abstract}
Traffic is inherently dangerous, with around 1.19 million fatalities annually. Automotive Mediated Reality (AMR) can enhance driving safety by overlaying critical information (e.g., outlines, icons, text) on key objects to improve awareness, altering objects' appearance to simplify traffic situations, and diminishing their appearance to minimize distractions. However, real-world AMR evaluation remains limited due to technical challenges.
To fill this \textit{sim-to-real} gap, we present \tool, an open-source tool that enables real-time AMR in real vehicles. \tool implements 15 effects across the AMR spectrum of augmented, diminished, and modified reality using state-of-the-art computational models for object detection and segmentation, depth estimation, and inpainting.
In an on-road expert user study (N=9) of \tool, participants enjoyed the AMR experience while pointing out technical limitations and identifying use cases for AMR. We discuss these results in relation to prior work and outline implications for AMR ethics and interaction design.
\end{abstract}



\begin{teaserfigure}
\centering
  \includegraphics[width=.8\linewidth]{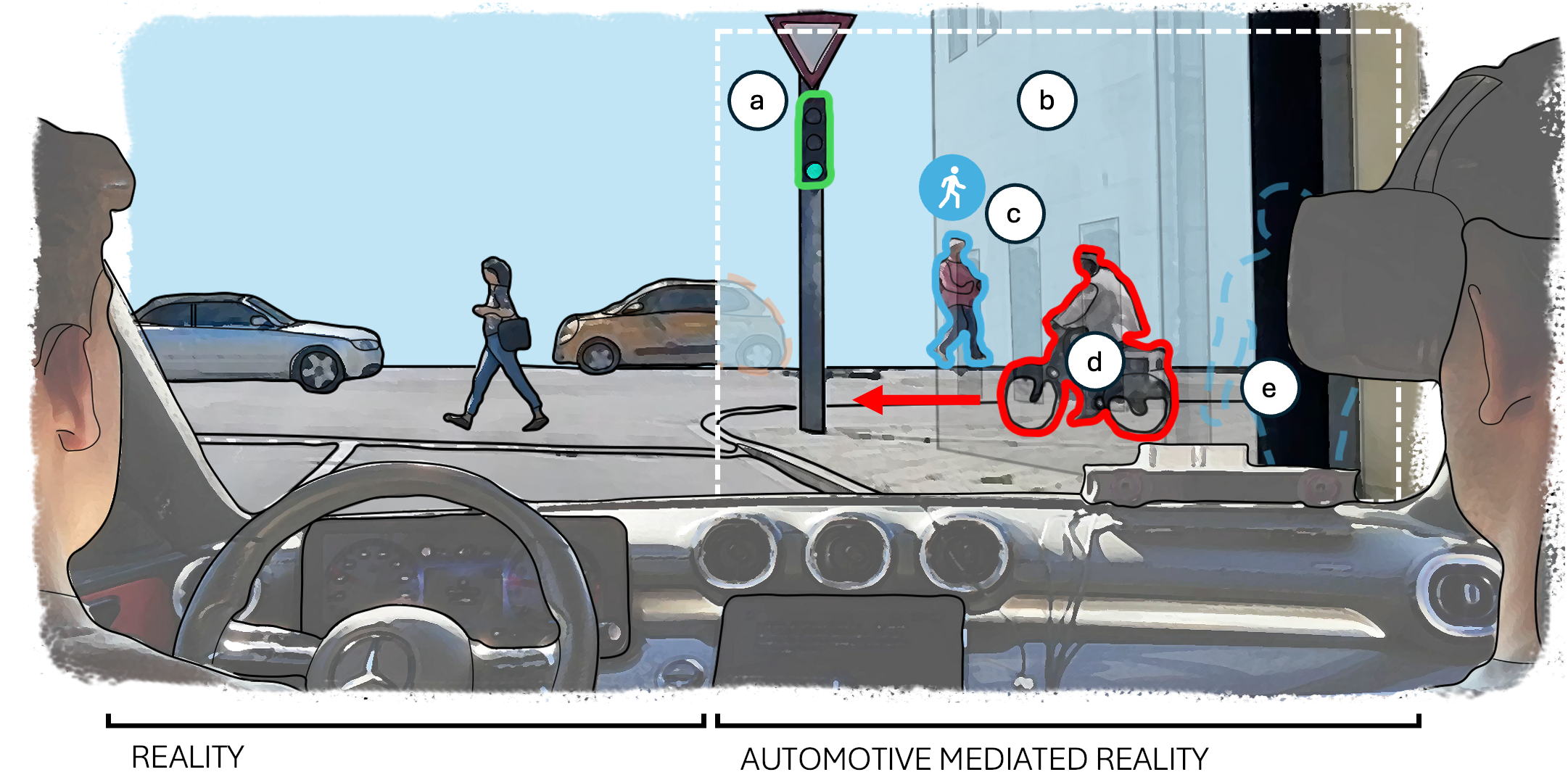}
  \caption{Concept of \tool enabling real-time Automotive Mediated Reality (AMR) in real vehicles. AMR alters driver or passenger perception by visually augmenting, diminishing, and modifying elements in the driving environment. For example: (a) highlighting traffic lights and road users, (b) diminishing the opacity of a building to reveal (c) a pedestrian with crossing intention, and (d) a cyclist about to enter the road; while another pedestrian is (e) diminished to reduce visual clutter. Such AMR concepts may support road safety or improve perceived safety, trust, and acceptance in complex traffic (e.g., \cite{muller_ar4cad_2022}), depending on the user's role (driver/passenger) and the level of driving automation. The image illustrates the prototyping setup used in our expert study: an expert user in the passenger seat wears an HMD running \tool, while an experimenter manually drives, as the current HMD-based prototype is not fully safe for driver use.}
  \Description{An illustration comparing reality (left) with an Automotive Mediated Reality (right). The viewer takes the perspective of a passenger in the backseat of a vehicle approaching an intersection. Two people are sitting in the front seats of the vehicle. The person on the left only perceives the reality: A person is crossing the street in front of the vehicle, and two cars are crossing the intersection. A traffic light on the sidewalk on the right displays a green light. The person on the right can view visualizations enabled through automotive mediated reality. The traffic light is highlighted with a green outline. A building on the right is made see-through by lowering its opacity, revealing vulnerable road users around the corner: A pedestrian is about to enter the road, and a cyclist is rapidly approaching from the right, likely to enter the road in front of the vehicle. Another pedestrian the vehicle is currently passing is walking on the sidewalk on the right. They are removed entirely to declutter the scene and reduce distractions.}
  \label{fig:teaser}
\end{teaserfigure}

\maketitle

\section{Introduction}
\label{ch:introduction}
According to the World Health Organisation's Global Status Report on Road Safety, around 1.19 million fatal road accidents occurred in 2023 alone~\cite{noauthor_global_nodate}. One reason is unsafe driving, often caused by inappropriate driver behavior, such as risky overtaking maneuvers and close following distances~\cite{destatis_fahrzeugfuehrer}. These behaviors are amplified by the rising complexity of mixed traffic \cite{rasouli2017understanding}, where the diversification of vulnerable road users, including pedestrians, bicycles, and e-scooters, increases interaction density, conflict potential, and the likelihood of driver error~\cite{EC_RoadSafety2024, ITF_SafeMicromobility}. The introduction of Automated Vehicles (AVs) further complicates traffic, as their behavior and reliability are challenging for manual drivers and AV passengers to assess~\cite{krome_how_2019} and may overwhelm them~\cite{theeuwes2017designing}.

To address these challenges, prior work~\cite{doula_ar-cp_2024, colley_effects_2022} has proposed vehicle interfaces that \textit{communicate} relevant traffic information to drivers and passengers (hereafter referred to as 'end-users'). This depends on the vehicle's level of automation. For example, at Society of Automotive Engineers (SAE)~\cite{noauthor_sae_nodate} levels 0–2, manual vehicles warn drivers of obstacles, while at levels 3–5, AVs communicate their state and decisions, such as detected road users and planned driving trajectories, to improve end-users' trust and perceived safety~\cite{jansen_visualizing_2024}, especially critical during the early stages of adopting AVs~\cite{kraus_more_2020}.

\textit{Situated} communication is essential, as traffic information depends on visual context. Previous work used Augmented Reality (AR) visualizations on Windshield Displays (WSDs) to superimpose relevant information onto the traffic situation. For instance, AR navigation cues improve drivers' Situation Awareness (SA)~\cite{doula_ar-cp_2024}, and visualizing the AV’s intended path on an AR WSD fosters end-user trust~\cite{colley_effects_2024}. However, traffic situations may also overwhelm end-users. To mitigate this, \citet{colley_feedback_2022} used Diminished Reality (DR) to remove information, such as vehicles in crowded intersections, in fully automated traffic. Moreover, using Modified Reality (ModR), entire environments or parts can be replaced (e.g., replacing an intense urban environment with a rural landscape to calm AV end-users \cite{colley_feedback_2022}). Collectively, AR, DR, and ModR approaches, which alter end-users’ perception of the physical world, can be defined as Mediated Reality (MR)~\cite{mann_mediated_1999}.

Most MR systems operate in controlled indoor scenes with stable geometry and bounded motion (e.g., Remixed Reality~\cite{10.1145/3173574.3173703}). These assumptions--static layouts, short viewing distances, and low risk--do not hold on the road. We thus propose \textit{Automotive Mediated Reality (AMR)} as a distinct design space for on-road AR, DR, and ModR.
Evaluating these in real vehicles is crucial for ecological validity, yet real-world studies require costly equipment (e.g., LiDAR, cameras, and WSDs)~\cite{goedicke_xr-oom_2022, bu_portobello_2024}. Further, AMR approaches require precise object detection, segmentation, and filtering (e.g., by distance), which relies on computational models and powerful GPUs for real-time performance. Therefore, previous work has mostly relied on simulated environments, such as those using Virtual Reality (VR). Despite these challenges, prior work has already demonstrated AR~\cite{kim_what_2023, ghiurau_arcar_2020, bu_portobello_2024, schramm2025augmented}, DR~\cite{kari_transformr_2021, noauthorsamsungnodate}, and ModR~\cite{kari_transformr_2021, holoride, hock_carvr_2017} in real vehicles.

However, to our knowledge, no existing work integrates AR, DR, and ModR approaches into a single MR framework that enables combined visual augmentation, diminution, and modification of traffic situations perceived by end-users in real vehicles.
This \textit{sim-to-real} gap limits the replication of lab studies (e.g.,~\cite{colley_feedback_2022,jansen_visualizing_2024,colley_effects_2024}) and the development and evaluation of novel AMR approaches in real-world driving conditions. For instance, DR and ModR are little researched in real vehicles, although lab studies indicate they could reduce distractions and stress in busy traffic~\cite{colley_feedback_2022}.

To fill this gap, we introduce \textbf{M}ediation \textbf{I}nterface for \textbf{R}eal-time \textbf{A}utomotive \textbf{G}raphics and \textbf{E}ffects (\tool), an open-source \href{https://unity.com/}{Unity}-based tool enabling real-time AMR in any vehicle, see \autoref{fig:teaser}.
\tool creates a WSD using video passthrough from a Head-Mounted Display (HMD). An RGB camera on the vehicle captures the environment and projects it to the WSD. The camera input is processed through a modular pipeline featuring state-of-the-art computational models. A semantic segmentation model~\cite{yolo11_ultralytics} classifies the environment into objects (e.g., cars, trees, and buildings), while a zero-shot depth estimation model~\cite{depth_anything_v2} calculates object distances. Lastly, \textit{AMR effects} are applied: graphical modifications to these objects and visualizations within the driving environment.

We demonstrate \tool with 15 AMR effects grouped into three categories:
(1) AR effects include highlighting object outlines, displaying additional information (e.g., icons and text attached to objects or freely placed in the environment), and marking objects with bounding boxes.
(2) DR effects contain removing objects via generative inpainting~\cite{sargsyan_migan_2023}, adjusting object transparency, blurring objects, and displaying only object outlines.
Finally, (3) ModR effects include replacing objects with other information, applying spatial transformations (translation, scaling, rotation), changing object colors, and artistic style transformations.

\tool is modular and runtime-customizable through a dedicated User Interface (UI), enabling integration of new computational models and AMR effects.
Models run directly in Unity with Unity Inference Engine~\cite{noauthorunitynodate}, and we implemented the effects with compute shaders and C\# scripts.

In this work, we distinguish between end-users (drivers and passengers who might experience AMR in future vehicles) and expert users (researchers and practitioners who may use \tool for prototyping). The current \tool prototype relies on an HMD with video passthrough and a UI panel for configuring AMR effects; therefore, it is not intended for use during manual driving. Instead, experts use the system in the passenger seat to prototype, combine, and critique AMR effects in situ. The AMR effect concepts, not the HMD-based tool itself, are what would target end-users.

To evaluate \tool, we deployed it in a real vehicle and conducted an expert user study (N=9) from industry (n=4) and academia (n=5). Participants created and experienced AMR during a 15-minute test drive. We gathered qualitative and quantitative feedback on the system, AMR effects, and the broader AMR concept.

\textit{Contribution Statement (following the typology of~\citet{Wobbrock.2016}):} 
\begingroup
\setlist[itemize]{itemsep=0pt, topsep=2pt} 
\begin{itemize}
    \item \textbf{Theoretical.} A categorization of Automotive Mediated Reality (AMR) effects, including concepts for augmenting, diminishing, and modifying driving environments.
    \item \textbf{Artifact or System.} \href{https://github.com/J-Britten/MIRAGE}{\tool}, an open-source tool for creating and evaluating real-time capable AMR effects in any vehicle.
    \item \textbf{Empirical study that tells us about how people use a system.} Insights from an expert user study (N=9) of \tool in a real vehicle, offering feedback on \tool, effects, and AMR. Based on this, we provide an outlook on the implications of ethics and interaction concepts for AMR.
\end{itemize}
\endgroup

\section{Related Work}
\label{ch:related-work}
\tool is grounded in research on (1) AMR approaches, (2) tools enabling research in real vehicles, and (3) computer vision models for real-time understanding of driving environments. 

\subsection{Mediated Reality Approaches in the Automotive Domain}
The term MR was introduced by \citet{mann_mediated_1999}, expanding Milgram’s reality-virtuality continuum~\cite{milgram_augmented_1995} by categorizing interactions with real-world objects more precisely. MR allows not only adding virtual elements to reality (AR) but also modifying (ModR) or removing existing elements (DR).
In recent years, automotive UIs increasingly feature HMDs and WSDs~\cite{jansen2022design}, enabling AR, DR, and ModR approaches in vehicles.

\textbf{AR} approaches, such as visualizations on WSDs, significantly enhance driver assistance systems by providing faster and more intuitive information compared to traditional displays~\cite{bauerfeind_navigating_2021}. They enhance driver performance in critical tasks~\cite{calvi_effectiveness_2020,calvi_evaluation_2020}, improve drivers' SA during SAE level 3 vehicle takeovers~\cite{lindemann_catch_2018,doula_ar-cp_2024,shi_effects_2024}, and increase end-user trust in SAE levels 3-4 through trajectory visualization and object highlighting~\cite{colley_effects_2024,wintersberger_explainable_2020,colley_effects_2022}. AR also helps communicate automation uncertainty, further enhancing trust in AVs~\cite{doula_can_2023,kunze_augmented_2018,von_sawitzky_increasing_2019,frison_ux_2019,kraus_more_2020} and has been explored for infotainment~\cite{wang_reducing_2019,berger_ar-enabled_2021,kim_-vehicle_2025}.
However, AR is typically researched in isolation rather than combined with, for example, DR and ModR. Yet, AR alone could cause information overload in busy traffic~\cite{jansen2025opticarvis}.

\textbf{DR} approaches visually reduce or remove objects from a end-user’s view~\cite{mori_survey_2017}, supporting decluttering~\cite{chan_declutterar_2022,yokoro_decluttar_2023,pintore_instant_2022}, reducing distractions~\cite{kim_dont_2020,katins_ad-blocked_2025}, or making objects transparent~\cite{kalkofen_adaptive_2013,cheng_towards_2022}. DR effects include opacity reduction, blurring, and desaturation~\cite{cheng_towards_2022}. Automotive DR uses exterior cameras and vehicle-to-everything communication to visualize blind spots~\cite{lindemann_diminished_2017,rameau_real-time_2016,maruta_blind-spot_2021}, as demonstrated by Samsung’s truck-mounted overtaking aid~\cite{noauthorsamsungnodate}. DR also mitigates sun glare~\cite{hong_visual_2024} and motion sickness~\cite{sawabe_diminished_2016}. However, DR may unintentionally hide critical information~\cite{kim_ciro_2021,kim_dont_2020,colley_feedback_2022}, emphasizing the importance of real-world evaluation, which remains underexplored.

Instead of reducing information, \textbf{ModR} approaches modify objects or environments, such as replacing objects, applying transformations, or changing visual styles. For instance, \citet{kari_transformr_2021} replaced vehicles with abstract or animal forms. Commercial implementations like \citet{holoride} and research \cite{hock_carvr_2017,stampf_move_2024,holoride} transform the vehicle’s surroundings into virtual environments synchronized with actual driving movements. Additionally, ModR can employ artistic style transfer to provide less realistic yet visually engaging representations of environments or objects~\cite{castillo_son_2017, kurzman_class-based_2019, lin_image_2021}.

Previous work often evaluated AR, DR, and ModR in controlled simulator scenarios. This raises uncertainty about their effectiveness across diverse real-world contexts (e.g., urban vs. rural environments)~\cite{flohr_value_2023}. Moreover, in SAE levels 3-4, end-users often lack understanding of situation-dependent AV behaviors, affecting their trust, perceived safety, and SA~\cite{jansen_visualizing_2024}. Thus, evaluating AMR approaches in real, diverse driving scenarios is essential.

Therefore, we provide a holistic perspective on AMR (i.e., considering AR, DR, and ModR). \tool enables expert users to create, customize, and evaluate AMR approaches in real vehicles. Unlike prior work, the \tool concept enables a flexible combination of AMR effects. For instance, irrelevant visual distractions like billboards can be reduced in DR, while AR highlights crossing pedestrians behind an obstructed road bend.

\subsection{On-Road Mediated Reality Research Tools}
Traditionally, automotive UI evaluations relied on simulation frameworks such as \textit{CARLA}~\cite{carla2017}, \textit{AirSim}~\cite{airsim2017}, and \textit{DriveSimQuest}~\cite{chidambaram2025drivesimquest}. These platforms provide safe, controlled, repeatable, and immersive VR HMD-based experimentation but cannot fully replicate the perceptual dynamics, motion cues, and situational variability of real vehicles. As a result, they leave a sim-to-real gap that AMR research must ultimately bridge.

HMD-based research tools deployed in moving vehicles may address this gap~\cite{argui_advancements_2024}. Prior work has demonstrated HMD setups in real cars~\cite{kim_what_2023, schramm2025augmented, sasalovici_bumpy_2025} and for immersive in-vehicle data analysis~\cite{jansen_autovis_2023}.
Furthermore, prior work used HMDs paired with 360° external cameras~\cite{yeo_toward_2020, kim_what_2023}, placing participants virtually in the driver’s seat. \citet{bu_portobello_2024} presented \textit{Portobello}, enabling lab-to-real studies based on \textit{XR-OOM}~\cite{goedicke_xr-oom_2022}. However, these methods rely on expensive, complex 3D LiDAR setups, limiting scalability. Conversely, \textit{PassengXR}~\cite{mcgill_passengxr_2022} offers a low-cost solution using Arduino sensors with mobile HMDs (e.g., Meta Quest), but suffers from limited computing capabilities. Moreover, these tools are limited to AR and some ModR effects (i.e., only replacing the entire driving environment with a virtual environment similar to \citet{holoride}).

To address these limitations, we introduce \tool, requiring only a computer, HMD, and a camera mounted to the vehicle's front exterior. Unlike prior tools, \tool supports AMR holistically, enabling real-vehicle DR and ModR besides AR while reducing technical complexity. Thus, expert users can more easily replicate simulator studies and develop and evaluate novel AMR approaches, for example, allowing earlier identification of conceptual issues in real driving environments.

\subsection{Computer Vision Models to Understand Driving Environments}
In lab simulations (e.g., in Unity), expert users have complete control and immediate access to environmental data. In real-world scenarios, however, this information must first be captured and processed using computer vision models. Object detection and semantic segmentation models like \textit{YOLO}~\cite{Jocher_Ultralytics_YOLO_2023} classify and identify scene objects, enabling object-based AMR effects. Additionally, inpainting models such as \textit{Lama}~\cite{suvorov2021resolution} or \textit{MI-GAN}~\cite{sargsyan_migan_2023} allow DR and ModR effects by removing and replacing objects.

Recent advances enable segmentation and inpainting models to run in real-time (up to 20 ms inference) on consumer-grade hardware. For instance, \textit{TransforMR}~\cite{kari_transformr_2021} achieves real-time DR through segmentation, inpainting, and 3D object substitution but requires powerful cloud-based infrastructure (e.g., four NVIDIA Tesla V100 GPUs~\cite{kari_transformr_2021}), increasing complexity, cost, and latency~\cite{arthurs_taxonomy_2022}. \citet{kim_what_2023} presented a segmentation approach in real vehicles running on local hardware in the vehicle trunk, displaying segmented areas on a virtual windshield via Python-Unity communication. However, this approach only supports AR effects.

We extend the in-vehicle computation approach of \citet{kim_what_2023} by integrating models directly into Unity using the Unity Inference Engine~\cite{noauthorunitynodate} to simplify the software setup. \tool incorporates pipelines for object identification, segmentation (\textit{YOLO11}~\cite{yolo11_ultralytics}), inpainting (\textit{MI-GAN}), and depth estimation (\textit{DepthAnythingV2}), achieving up to 34 FPS, enabling real-time AMR.



\begin{figure*}[ht]
    \centering
    \includegraphics[width=\linewidth]{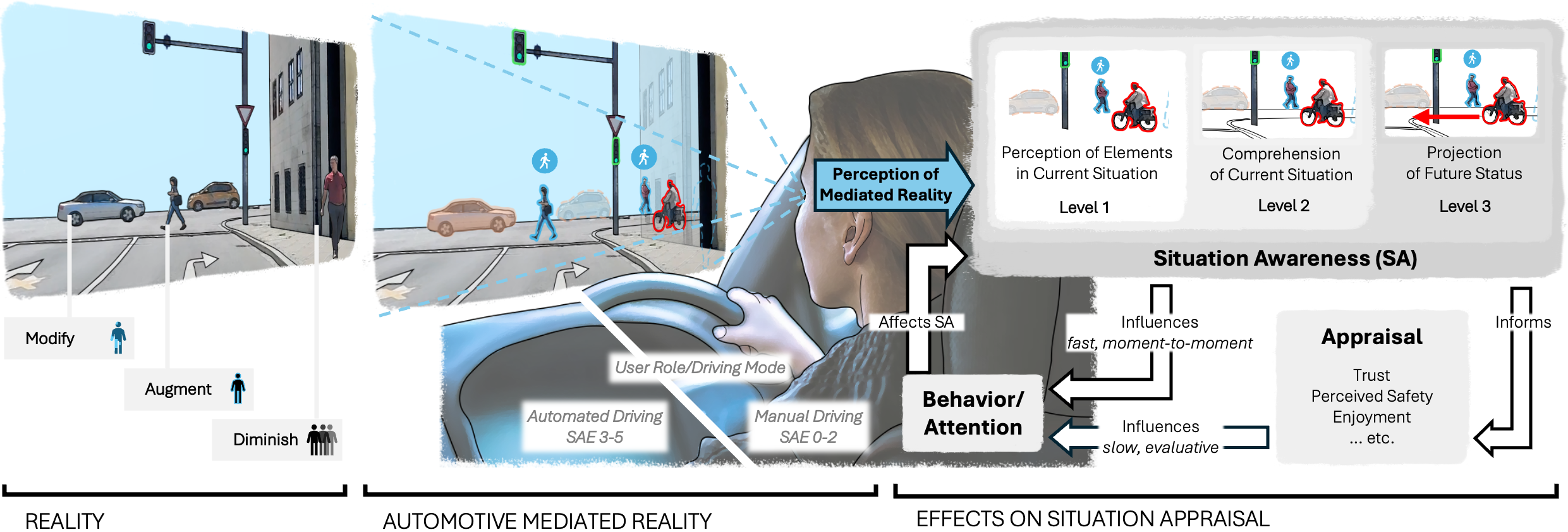}
    \caption{Automotive Situation Awareness–Appraisal Mediation Cycle. AMR effects (Modify, Augment, Diminish) alter the visual scene before perception, shaping SA across Levels 1–3~\cite{endsley2017toward}. SA influences behavior and attention in a fast, moment-to-moment manner~\cite{endsley2017toward}, whereas downstream appraisal (e.g., trust~\cite{lee2004trust}, perceived safety~\cite{carsten2019can}, enjoyment in non-driving tasks~\cite{hock_carvr_2017,schramm2025augmented}) shapes behavior and attention more slowly and evaluatively~\cite{scherer2001appraisal}. Appraisal is further informed by expectations, prior experience, goals, and affective states~\cite{scherer2001appraisal}, although these influences are not explicitly visualized in the cycle. Behavior and attention affect subsequent SA~\cite{endsley2017toward}. The mediation cycle applies across automation levels (drivers in SAE 0–2 manual driving; passengers in SAE 3–5 automated driving), where AMR may support different driving or non-driving tasks.}
    \label{fig:situation-appraisal-mediation}
    \Description{A figure depicting the automotive situation awareness-appraisal cycle. It is split into three parts. On the left, the scene of an intersection is depicted with cars and pedestrian. In the center, the same scene is depicted, but mediated: Cars are painted in different colors, pedestrians have outlines and a building on the right is made transparent to reveal a cyclist behind the corner. A woman in a car perceives this AMR, which is the starting point for the third section: the effects on situation appraisal cycle. The perception AMR affects the three levels of situation awareness, which in turn affects behavior and attention, but also appraisal (trust, perceived safety, enjoyment) which then also influences behavior or attention. This ultimately affects SA again. The figure depicts the cockpit of the car in a split view with and without a steering wheel to highlight the relevance for AMR for manual and automated driving.}
\end{figure*}

\section{Conceptual Framing of Automotive Mediated Reality}
\label{ch:concept}
We conceptualize AMR as a layer in a perceptual process that mediates visual information before it reaches the end-user (see \autoref{fig:situation-appraisal-mediation}). These mediations influence two well-established cognitive constructs central to interaction in vehicles \cite{carsten2019can}: SA and appraisal. Our framing builds on Mann’s definition of MR~\cite{mann_mediated_1999}, which introduced augmentation, diminution, and modification of visual input across general-purpose contexts such as wearable computing and assistive vision. It also differs from MR frameworks such as \citet{10.1145/3173574.3173703}, which focus on indoor scenes with stable geometry and predictable motion patterns. In contrast, the automotive context imposes safety-critical perceptual demands: end-users must maintain SA to perceive relevant elements, comprehend their meaning, and anticipate future developments in a dynamic, high-risk environment \cite{theeuwes2017designing}. They also form appraisals such as trust in automation\cite{lee2004trust}, perceived safety~\cite{carsten2019can}, comfort~\cite{domova2024comfort}, and enjoyment during non-driving tasks~\cite{hock_carvr_2017}, which influence their behavior and attention when interacting with vehicle systems (e.g., driving assistance~\cite{frison_ux_2019}).

In the following sections, we detail this framing and outline the AMR effects, as well as the design space that informed the implementation of \tool.

\begin{figure*}[ht]
    \centering
    \includegraphics[width=1\linewidth]{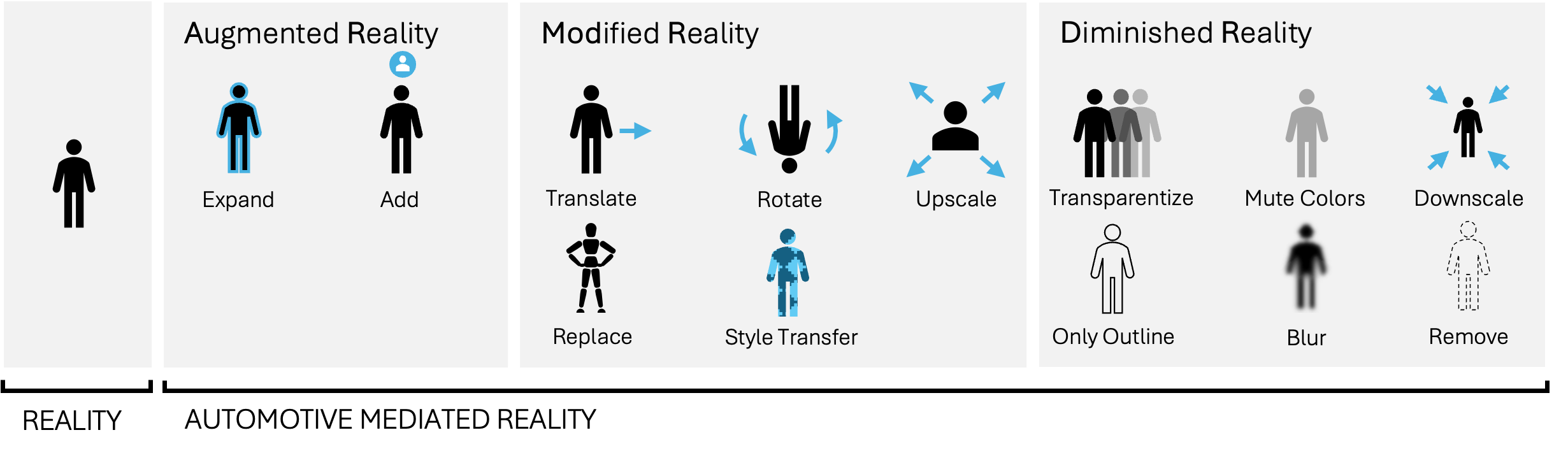}
    \caption{AMR includes AR, DR, and ModR approaches. AR adds information to the driving environment and expands information on specific objects. DR reduces the visual information of objects, making them less detailed. ModR modifies objects by translating, rotating, scaling, changing their art style, and replacing them entirely.}
    \label{fig:mr-concept}
    \Description{A horizontal diagram compares reality (left) with various Automotive Mediated Reality (AMR) effects (right). It is split into 4 boxes: The first depicts reality. Next, two icons represent Augmented Reality (labeled “Add” and “Expand”).  After that, Modified Reality (showing icons for translating, rotating, and enlarging objects, replacing them with new graphics, or applying a stylized appearance) and Diminished Reality (showing icons for reducing opacity (transparentizing), color, and scale, outlining only, blurring, or removing elements). Silhouettes illustrate each effect category and its impact on a basic human figure.}
\end{figure*}

\subsection{Automotive Situation Awareness–Appraisal Mediation Cycle}
\label{sec:mediation-cycle}
AMR inserts a mediation layer between the physical driving environment and the end-user’s perception (see \autoref{fig:situation-appraisal-mediation}). Because SA relies on visually perceived information, changes introduced by AMR affect SA at the point of perceptual input.
According to Endsley~\cite{endsley2017toward}, SA comprises three levels: L1 \textit{Perception} (detecting relevant elements, e.g., vehicles, pedestrians, signs), L2 \textit{Comprehension} (understanding relations and situational meaning, e.g., a pedestrian waiting to cross indicating potential hazard), and L3 \textit{Projection} (anticipating future states, e.g., predicting that the pedestrian may enter the road).
These levels form interdependent layers of situational knowledge rather than strict processing stages \cite{endsley2017toward}. AMR may therefore affect them differently.
For example, at L1, AR highlighting may increase road user visibility; at L2, the same highlighting may increase visual clutter, complicating relational interpretation; and at L3, highlighting motion-relevant elements may shift attention toward cues used to project future behavior.

SA supplies interpreted situational information \cite{endsley2017toward}, which forms part of the input to appraisal processes. Appraisal refers to cognitive–affective evaluations of a situation \cite{scherer2001appraisal} and can manifest in constructs such as trust \cite{lee2004trust}, perceived safety \cite{carsten2019can}, comfort \cite{domova2024comfort}, and hedonic responses relevant to non-driving tasks (e.g., enjoyment of sightseeing overlays or entertainment content \cite{hock_carvr_2017, schramm2025augmented}). These evaluations integrate the interpreted situation with expectations, prior experience, goals, and affective states \cite{scherer2001appraisal}. Consequently, AMR influences appraisal indirectly by affecting SA.

SA and appraisal influence behavior and attention in different ways. SA influences fast, moment-to-moment behavior and attention, such as hazard checking \cite{shi_effects_2024}. Appraisal influences slower, more deliberative behaviors such as reliance on automation \cite{lee2004trust} or attentional shifts \cite{carsten2019can}. Because attention affects which elements are perceived, behavioral and attentional changes influence subsequent SA~\cite{endsley2017toward}. Together, these processes form a mediation cycle.

Differences across automation levels primarily concern the SA required for control.
During manual driving (SAE 0–2), end-users need SA to support lane keeping, hazard anticipation, and speed control \cite{theeuwes2017designing}, while during supervisory monitoring (SAE 2–3), they need SA of automation behavior (e.g., lane centering, gap keeping) for an eventual takeover \cite{shi_effects_2024}.
During automated driving (SAE 3–5) and for passengers at all SAE levels, end-users perform non-driving tasks (e.g., sightseeing, contextual AR explanations, entertainment). Although SA demands for vehicle control decrease, appraisal remains important, as, for example, reduced perceived safety can hinder end-user acceptance of AVs in traffic \cite{nordhoff2018acceptance}.

\subsection{Automotive Mediated Reality Effects}
\label{sec:AMR-effects}
Following Mann’s definition of MR~\cite{mann_mediated_1999} and later MR work~\cite{10.1145/3173574.3173703}, we distinguish three categories of AMR effects (see \autoref{fig:mr-concept}):
\textbf{AR} effects add information to the driving environment and expand information on specific elements (e.g., adding predicted movement trajectories of road users \cite{colley_effects_2024}).
\textbf{ModR} effects modify the visual appearance of existing elements (e.g., adding visual filters using style transfers to increase their visibility \cite{colley2021effects}).
\textbf{DR} effects diminish visual information (e.g., removing distracting elements in the background \cite{colley_feedback_2022}).
AMR effects can also be combined, for instance, by removing non-critical elements via DR while simultaneously emphasizing others via AR highlighting.
Because AMR effects influence the visibility, interpretability, and predictability of elements associated with SA~L1–L3, they mediate the information on which end-users base appraisal and behavior/attention. Designing AMR applications, therefore, requires attention to how each mediation may support or interfere with SA and appraisal.

\begin{table*}[h]
\caption{Design Space for the conceptual exploration of AMR applications with two dimensions: AMR effects with example variants, and Information Type (Dynamic, Location-Specific, Action) and Content Reference Type (highlight, depict, combined) based on \textit{AR4CAD}~\cite{muller_ar4cad_2022}. While we organized the related work within the design space, it is not exhaustive, and empty cells do not necessarily imply research gaps.}
\Description{A table depicting the design space fo rthe conceptual exploration of AMR application with two dimensions: Along the columns, the Semantic Mediation Target along with Information Type (Dynamic, Location-Specific, Action) and Content Reference Type (Highlight, Depict, Combined) derived from AR4CAD (Muller et al.). Along the rows, AMR effect categories (augmented reality, modified reality, diminished reality) with example variants are depicted. The related work cited in this work was sorted into the cells, with most falling in the AR, Add row and the Action and Depict column. Generally, most works lie within the AR rows, with some works in the ModR and fewer in DR.}
\label{tab:designspace}
\resizebox{\ifdim\width>\linewidth\linewidth\else\width\fi}{!}{
\begin{tabular}{ccc|lllllllll|}
\cline{4-12}
\multicolumn{1}{l}{} & \multicolumn{1}{l}{} & \multicolumn{1}{l|}{} & \multicolumn{9}{c|}{\cellcolor[HTML]{F3F3F3}\textbf{Semantic Mediation Target}} \\ \cline{4-12} 
\multicolumn{1}{l}{} & \multicolumn{1}{l}{} & \multicolumn{1}{l|}{} & \multicolumn{3}{c|}{\cellcolor[HTML]{FFB7E4}Dynamic Environment Element} & \multicolumn{3}{c|}{\cellcolor[HTML]{FFB7E4}Location-Specific Element} & \multicolumn{3}{c|}{\cellcolor[HTML]{FFB7E4}Action Element} \\ \cline{4-12} 
\multicolumn{1}{l}{} & \multicolumn{1}{l}{} & \multicolumn{1}{l|}{} & \multicolumn{1}{c|}{\cellcolor[HTML]{FD6865}Highlight} & \multicolumn{1}{c|}{\cellcolor[HTML]{FD6865}Depict} & \multicolumn{1}{c|}{\cellcolor[HTML]{FD6865}Combined} & \multicolumn{1}{c|}{\cellcolor[HTML]{FD6865}Highlight} & \multicolumn{1}{c|}{\cellcolor[HTML]{FD6865}Depict} & \multicolumn{1}{c|}{\cellcolor[HTML]{FD6865}Combined} & \multicolumn{1}{c|}{\cellcolor[HTML]{FD6865}Highlight} & \multicolumn{1}{c|}{\cellcolor[HTML]{FD6865}Depict} & \multicolumn{1}{c|}{\cellcolor[HTML]{FD6865}Combined} \\ \hline
\rowcolor[HTML]{F3F3F3} 
\multicolumn{1}{|c|}{\cellcolor[HTML]{F3F3F3}} & \multicolumn{1}{c|}{\cellcolor[HTML]{F3F3F3}} & Expand & \multicolumn{1}{c|}{\cellcolor[HTML]{F3F3F3}\cite{lindemann_catch_2018, jansen2025opticarvis}} & \multicolumn{1}{l|}{\cellcolor[HTML]{F3F3F3}} & \multicolumn{1}{l|}{\cellcolor[HTML]{F3F3F3}} & \multicolumn{1}{l|}{\cellcolor[HTML]{F3F3F3}} & \multicolumn{1}{c|}{\cellcolor[HTML]{F3F3F3}\cite{kunze_augmented_2018}} & \multicolumn{1}{l|}{\cellcolor[HTML]{F3F3F3}} & \multicolumn{1}{l|}{\cellcolor[HTML]{F3F3F3}} & \multicolumn{1}{l|}{\cellcolor[HTML]{F3F3F3}} &  \\ \cline{3-3}
\multicolumn{1}{|c|}{\cellcolor[HTML]{F3F3F3}} & \multicolumn{1}{c|}{\multirow{-2}{*}{\cellcolor[HTML]{F3F3F3}AR}} & Add & \multicolumn{1}{c|}{\cite{colley_feedback_2022, kim_what_2023}} & \multicolumn{1}{c|}{\cite{doula_ar-cp_2024, colley_feedback_2022, 10.1145/3491101.3519741}} & \multicolumn{1}{c|}{\cite{calvi_effectiveness_2020, lindemann_catch_2018, colley_effects_2022, jansen2025opticarvis, manger_providing_2023}} & \multicolumn{1}{l|}{} & \multicolumn{1}{c|}{\cite{lindemann_catch_2018, kunze_augmented_2018, schramm2025augmented, manger_providing_2023, manger_explainability_2023, 10.1145/3491101.3519741}} & \multicolumn{1}{l|}{} & \multicolumn{1}{l|}{} & \multicolumn{1}{c|}{\begin{tabular}[c]{@{}c@{}}\cite{bauerfeind_navigating_2021, calvi_evaluation_2020, shi_effects_2024, colley_effects_2024, colley_effects_2022}\\ \cite{doula_can_2023, von_sawitzky_increasing_2019, wang_reducing_2019, berger_ar-enabled_2021,kim_-vehicle_2025}\\ \cite{jansen2025opticarvis, manger_providing_2023, manger_explainability_2023, 10.1145/3491101.3519741}\end{tabular}} & \multicolumn{1}{c|}{\cite{lindemann_catch_2018}} \\ \cline{2-3}
\rowcolor[HTML]{F3F3F3} 
\multicolumn{1}{|c|}{\cellcolor[HTML]{F3F3F3}} & \multicolumn{1}{c|}{\cellcolor[HTML]{F3F3F3}} & Translate & \multicolumn{1}{l|}{\cellcolor[HTML]{F3F3F3}} & \multicolumn{1}{l|}{\cellcolor[HTML]{F3F3F3}} & \multicolumn{1}{l|}{\cellcolor[HTML]{F3F3F3}} & \multicolumn{1}{l|}{\cellcolor[HTML]{F3F3F3}} & \multicolumn{1}{l|}{\cellcolor[HTML]{F3F3F3}} & \multicolumn{1}{l|}{\cellcolor[HTML]{F3F3F3}} & \multicolumn{1}{l|}{\cellcolor[HTML]{F3F3F3}} & \multicolumn{1}{l|}{\cellcolor[HTML]{F3F3F3}} &  \\ \cline{3-3}
\multicolumn{1}{|c|}{\cellcolor[HTML]{F3F3F3}} & \multicolumn{1}{c|}{\cellcolor[HTML]{F3F3F3}} & Rotate & \multicolumn{1}{l|}{} & \multicolumn{1}{l|}{} & \multicolumn{1}{l|}{} & \multicolumn{1}{l|}{} & \multicolumn{1}{l|}{} & \multicolumn{1}{l|}{} & \multicolumn{1}{l|}{} & \multicolumn{1}{l|}{} &  \\ \cline{3-3}
\rowcolor[HTML]{F3F3F3} 
\multicolumn{1}{|c|}{\cellcolor[HTML]{F3F3F3}} & \multicolumn{1}{c|}{\cellcolor[HTML]{F3F3F3}} & Upscale & \multicolumn{1}{l|}{\cellcolor[HTML]{F3F3F3}} & \multicolumn{1}{l|}{\cellcolor[HTML]{F3F3F3}} & \multicolumn{1}{l|}{\cellcolor[HTML]{F3F3F3}} & \multicolumn{1}{l|}{\cellcolor[HTML]{F3F3F3}} & \multicolumn{1}{l|}{\cellcolor[HTML]{F3F3F3}} & \multicolumn{1}{l|}{\cellcolor[HTML]{F3F3F3}} & \multicolumn{1}{l|}{\cellcolor[HTML]{F3F3F3}} & \multicolumn{1}{l|}{\cellcolor[HTML]{F3F3F3}} &  \\ \cline{3-3}
\multicolumn{1}{|c|}{\cellcolor[HTML]{F3F3F3}} & \multicolumn{1}{c|}{\cellcolor[HTML]{F3F3F3}} & Replace & \multicolumn{1}{l|}{} & \multicolumn{1}{c|}{\cite{colley_feedback_2022, kari_transformr_2021, holoride, hock_carvr_2017, stampf_move_2024, 10.1145/3491101.3519741}} & \multicolumn{1}{c|}{\cite{doula_ar-cp_2024}} & \multicolumn{1}{l|}{} & \multicolumn{1}{c|}{\cite{sawabe_diminished_2016, colley_feedback_2022, holoride, hock_carvr_2017, stampf_move_2024}} & \multicolumn{1}{l|}{} & \multicolumn{1}{l|}{} & \multicolumn{1}{l|}{} &  \\ \cline{3-3}
\rowcolor[HTML]{F3F3F3} 
\multicolumn{1}{|c|}{\cellcolor[HTML]{F3F3F3}} & \multicolumn{1}{c|}{\multirow{-5}{*}{\cellcolor[HTML]{F3F3F3}ModR}} & Style Transfer & \multicolumn{1}{c|}{\cellcolor[HTML]{F3F3F3}\cite{colley_effects_2022, jansen2025opticarvis, kurzman_class-based_2019, jansen_visualizing_2024}} & \multicolumn{1}{l|}{\cellcolor[HTML]{F3F3F3}} & \multicolumn{1}{l|}{\cellcolor[HTML]{F3F3F3}} & \multicolumn{1}{c|}{\cellcolor[HTML]{F3F3F3}\cite{kurzman_class-based_2019}} & \multicolumn{1}{l|}{\cellcolor[HTML]{F3F3F3}} & \multicolumn{1}{l|}{\cellcolor[HTML]{F3F3F3}} & \multicolumn{1}{l|}{\cellcolor[HTML]{F3F3F3}} & \multicolumn{1}{l|}{\cellcolor[HTML]{F3F3F3}} &  \\ \cline{2-3}
\multicolumn{1}{|c|}{\cellcolor[HTML]{F3F3F3}} & \multicolumn{1}{c|}{\cellcolor[HTML]{F3F3F3}} & Transparentize & \multicolumn{1}{l|}{} & \multicolumn{1}{l|}{} & \multicolumn{1}{l|}{} & \multicolumn{1}{l|}{} & \multicolumn{1}{c|}{\cite{lindemann_diminished_2017, lindemann_examining_2017, lindemann_acceptance_2019}} & \multicolumn{1}{l|}{} & \multicolumn{1}{l|}{} & \multicolumn{1}{l|}{} &  \\ \cline{3-3}
\rowcolor[HTML]{F3F3F3} 
\multicolumn{1}{|c|}{\cellcolor[HTML]{F3F3F3}} & \multicolumn{1}{c|}{\cellcolor[HTML]{F3F3F3}} & Mute Colors & \multicolumn{1}{l|}{\cellcolor[HTML]{F3F3F3}} & \multicolumn{1}{l|}{\cellcolor[HTML]{F3F3F3}} & \multicolumn{1}{l|}{\cellcolor[HTML]{F3F3F3}} & \multicolumn{1}{l|}{\cellcolor[HTML]{F3F3F3}} & \multicolumn{1}{l|}{\cellcolor[HTML]{F3F3F3}} & \multicolumn{1}{l|}{\cellcolor[HTML]{F3F3F3}} & \multicolumn{1}{l|}{\cellcolor[HTML]{F3F3F3}} & \multicolumn{1}{l|}{\cellcolor[HTML]{F3F3F3}} &  \\ \cline{3-3}
\multicolumn{1}{|c|}{\cellcolor[HTML]{F3F3F3}} & \multicolumn{1}{c|}{\cellcolor[HTML]{F3F3F3}} & Downscale & \multicolumn{1}{l|}{} & \multicolumn{1}{l|}{} & \multicolumn{1}{l|}{} & \multicolumn{1}{l|}{} & \multicolumn{1}{l|}{} & \multicolumn{1}{l|}{} & \multicolumn{1}{l|}{} & \multicolumn{1}{l|}{} &  \\ \cline{3-3}
\rowcolor[HTML]{F3F3F3} 
\multicolumn{1}{|c|}{\cellcolor[HTML]{F3F3F3}} & \multicolumn{1}{c|}{\cellcolor[HTML]{F3F3F3}} & Only Outline & \multicolumn{1}{l|}{\cellcolor[HTML]{F3F3F3}} & \multicolumn{1}{l|}{\cellcolor[HTML]{F3F3F3}} & \multicolumn{1}{l|}{\cellcolor[HTML]{F3F3F3}} & \multicolumn{1}{l|}{\cellcolor[HTML]{F3F3F3}} & \multicolumn{1}{l|}{\cellcolor[HTML]{F3F3F3}} & \multicolumn{1}{l|}{\cellcolor[HTML]{F3F3F3}} & \multicolumn{1}{l|}{\cellcolor[HTML]{F3F3F3}} & \multicolumn{1}{l|}{\cellcolor[HTML]{F3F3F3}} &  \\ \cline{3-3}
\multicolumn{1}{|c|}{\cellcolor[HTML]{F3F3F3}} & \multicolumn{1}{c|}{\cellcolor[HTML]{F3F3F3}} & Blur & \multicolumn{1}{l|}{} & \multicolumn{1}{l|}{} & \multicolumn{1}{l|}{} & \multicolumn{1}{l|}{} & \multicolumn{1}{l|}{} & \multicolumn{1}{l|}{} & \multicolumn{1}{l|}{} & \multicolumn{1}{l|}{} &  \\ \cline{3-3}
\rowcolor[HTML]{F3F3F3} 
\multicolumn{1}{|c|}{\multirow{-13}{*}{\cellcolor[HTML]{F3F3F3}\begin{tabular}[c]{@{}c@{}}\textbf{AMR}\\ \textbf{Effects}\end{tabular}}} & \multicolumn{1}{c|}{\multirow{-6}{*}{\cellcolor[HTML]{F3F3F3}DR}} & Remove & \multicolumn{1}{l|}{\cellcolor[HTML]{F3F3F3}} & \multicolumn{1}{c|}{\cellcolor[HTML]{F3F3F3}\cite{rameau_real-time_2016, noauthorsamsungnodate, colley_feedback_2022}} & \multicolumn{1}{l|}{\cellcolor[HTML]{F3F3F3}} & \multicolumn{1}{l|}{\cellcolor[HTML]{F3F3F3}} & \multicolumn{1}{c|}{\cellcolor[HTML]{F3F3F3}\cite{lindemann_diminished_2017, lindemann_examining_2017, maruta_blind-spot_2021, colley_feedback_2022, lindemann_acceptance_2019}} & \multicolumn{1}{l|}{\cellcolor[HTML]{F3F3F3}} & \multicolumn{1}{l|}{\cellcolor[HTML]{F3F3F3}} & \multicolumn{1}{l|}{\cellcolor[HTML]{F3F3F3}} &  \\ \hline
\end{tabular}}
\end{table*}

\subsection{Design Space for Automotive Mediated Reality}
\label{sec:design-space}
We introduce a two-dimensional design space intended for the conceptual exploration of AMR applications (see \autoref{tab:designspace}). We employ a morphological matrix, also known as a Zwicky Box \cite{zwicky_MorphologicalApproachDiscovery_1967}, commonly used in HCI design-space research (e.g., \cite{jansen2022design}).
The first dimension comprises AMR effects with example variants categorized into AR, ModR, and DR (see \autoref{fig:mr-concept}).
The second dimension is based on the \textit{AR4CAD} taxonomy~\cite{muller_ar4cad_2022}. This taxonomy originally differentiates four types: Content Reference Type, Information Type, Functionality Type, and Registration Type, which describe the information communicated and its relationship to the environment developed for AR interfaces in AVs. 
We incorporate two types:
(1) Content Reference Type, distinguishing whether the mediation \textit{highlights} existing physical elements, \textit{depicts} information not present in the scene, or \textit{combines} both.
(2) Information Type, classifying the object of communication into \textit{dynamic environment elements} (e.g., vehicles, pedestrians), \textit{location-specific elements} (e.g., infrastructure, signs), and \textit{action elements} (e.g., predicted movement trajectories).
Crossing these types yields 9 semantic mediation targets that describe what an AMR effect mediates (see \autoref{tab:designspace}).

Our design space focuses on characterizing the information being mediated and how it relates to the environment, rather than prescribing how end-users should respond to it.
Accordingly, we exclude \textit{AR4CAD}'s Functionality Type (instructive vs. declarative) as it captures the intended behavioral effect of an AMR effect—whether it should guide a driver’s action or inform a passenger.
We also exclude the Registration Type (contact-analog, 2D, angle-analog, unregistered), as it concerns spatial anchoring and rendering technique rather than semantic mediation.
However, both types can be additional orthogonal dimensions in the design space.

While the design space is not a taxonomy and does not imply research gaps, consistent with morphological analysis, it provides a conceptual scaffold that helps pair AMR effects with semantic mediation targets. For example, \textit{highlighting} dynamic elements (e.g., vehicles \cite{wintersberger_explainable_2020, colley_effects_2022}) may support SA L1, \textit{depicting} action elements (e.g., predicted vehicle trajectories \cite{colley_effects_2024}) may support SA L3, and \textit{combining} depictive and highlight cues (e.g., highlighting a pedestrian while simultaneously displaying a predicted crossing path \cite{kim2016virtual}) may improve or decrease trust depending on whether these cues remain temporally and spatially coherent. Entertainment-oriented AMR (e.g., sightseeing overlays \cite{schramm2025augmented}) fits within \textit{highlighting}, \textit{depicting}, or \textit{combined} mediation of non-driving information.

\subsection{Tool Design Objectives}
\label{sec:objectives}
While AMR effects for all SAE levels have been explored in prior works (e.g.,~\cite{bauerfeind_navigating_2021, manger_providing_2023, colley_effects_2024, wintersberger_explainable_2020, colley_effects_2022, frison_ux_2019, kraus_more_2020}), their evaluations were limited to lab-based simulators, likely due to technical challenges in real-vehicle experiments. While simulations readily provide object data for visual modifications (e.g., color, outlines, removal), real-world setups must first identify objects and parameters (e.g., shape or distance) in real time, significantly increasing complexity and cost.
Based on these limitations, we identified four design objectives:

\begingroup
\setlist[enumerate]{itemsep=0pt, topsep=2pt} 
\begin{enumerate}[noitemsep]
    \itema \textbf{Holisticity}: \tool should support the whole spectrum of AMR, including AR, DR, and ModR.
    \itemb \textbf{Modularity}: AMR effects should be modular and combinable, for instance, as AR may introduce informational overload (e.g., visual outlines highlighting all nearby road users), while DR may irritate end-users when removed elements cause visual inconsistencies (e.g., an occluded cyclist suddenly disappearing behind an inpainted vehicle). ModR can similarly cause irritation when modified elements distort expected scene semantics (e.g., a vehicle substituted with an implausible background pattern). Additionally, \tool should be easy to use and modify, enabling expert users to add custom AMR effects.
    \itemc \textbf{Real-Time Performance}: To appropriately test AMR in real vehicles, \tool must detect, identify, and segment objects, as well as apply AMR effects in real-time.
    \itemd \textbf{Minimal Hardware Setup and Scalability}: To simplify deploying AMR in real vehicles, \tool should require as little hardware as possible.
\end{enumerate}
\endgroup

\begin{figure*}[ht]
    \centering
    \includegraphics[width=1\linewidth]{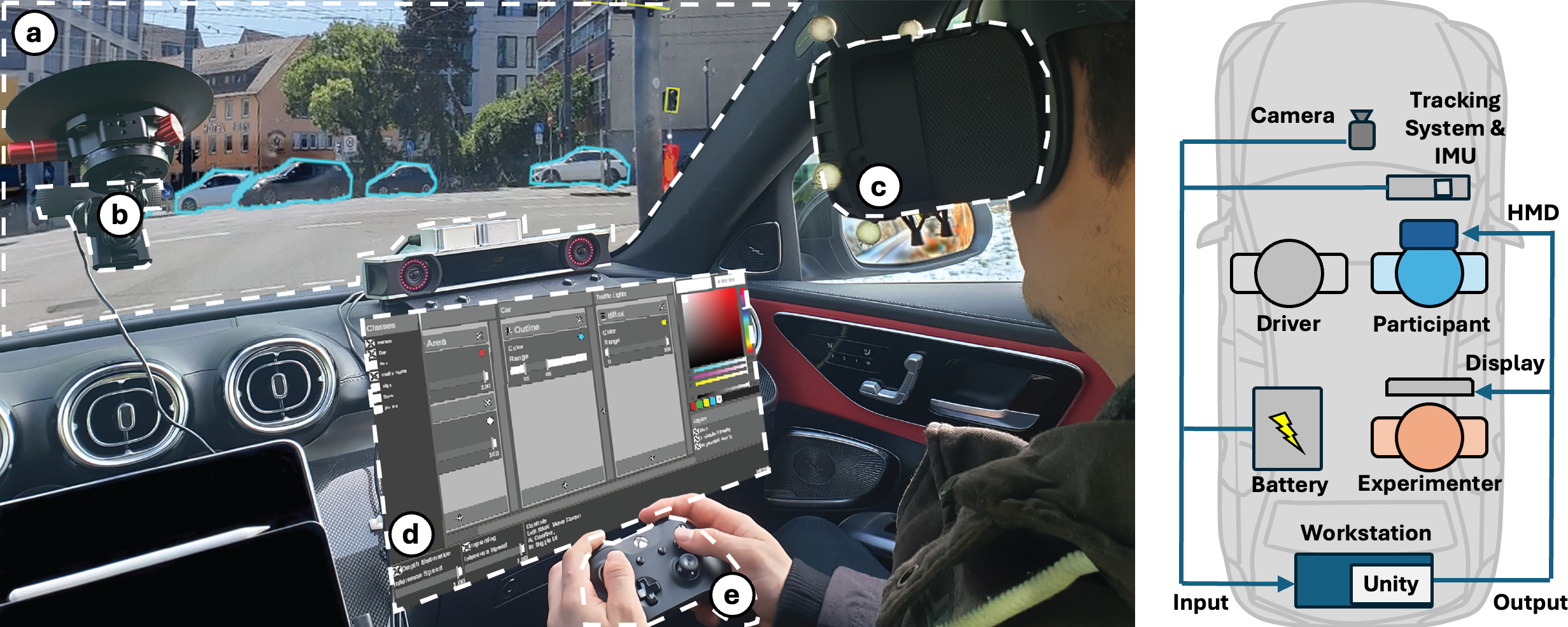}
    \caption{Overview of \tool. (a) A virtual WSD displays AMR effects. In this example, a blue outline is added to cars, pedestrians are colored red with icons above their heads, and traffic lights have yellow bounding boxes. The input image for the pipeline is recorded by a forward-facing camera (b). The user sees the WSD through an HMD (c). AMR effects can be controlled via UI (d) using a gamepad (e). The driver and experimenter were present during the expert user study (see Section \ref{ch:study}).}
    \label{fig:mirage_overview}
    \Description{A combined image shows an in-vehicle setup on the left and a schematic diagram on the right. On the left, a windshield-mounted camera (labeled “a”) captures the outside environment, highlighted with outlines around cars; a head-mounted display (labeled “c”) worn by the participant provides a simulated windshield view; and a controller (labeled “e”) manages AMR effects. A laptop interface (labeled “d”) runs \tool. On the right, a top-down schematic depicts the car interior, illustrating the hardware flow: the camera feeds into a workstation running Unity, which outputs to the HMD; additional components include a tracking system and IMU, a battery power source, and roles for driver, participant, and experimenter.}
\end{figure*}

\section{\tool: Implementation}
\label{sec:implementation}
\tool requires an RGB camera, an HMD with passthrough, and a powerful GPU; see \autoref{fig:mirage_overview}. We used an \textit{Nvidia RTX 3090 TI} in our study, but conducted benchmarks on a variety of other GPUs (see Section \ref{ssec:performance}). \tool is kept as simple as possible by running entirely within Unity v6000.2.2f1, eliminating the need for cloud computing or external AI inference (see \citet{kim_what_2023} and \citet{kari_transformr_2021}). Besides the HMD mode, \tool also offers a desktop mode that allows effects to be quickly tested on pre-recorded videos.

\subsection{Windshield Display Simulation}
\label{ssec:display}
One ideal way of enabling AMR would be WSDs, as HMDs add strain. However, as WSDs are not readily available, we created a virtual WSD using an RGB camera mounted to the windshield or hood, ideally with a wide-angle lens covering the entire windshield. Live camera input is projected onto a curved virtual surface, scaled to match real-world dimensions, creating the illusion of a WSD. We use stencil testing~\cite{noauthorlearnopenglnodate} with a windshield-shaped virtual plane to limit visuals strictly to the windshield area, avoiding depth artifacts within the vehicle. Extending to side windows is available as a setting in \tool with a broader camera field of view.

The virtual WSD is viewed through an HMD with passthrough functionality, enabling end-users to see the real environment overlaid with virtual AMR effects. We used a \href{https://varjo.com/products/varjo-xr-3/}{Varjo XR 3} in the expert evaluation and tested with a \href{https://www.meta.com/quest/quest-3/}{Meta Quest 3} during development.

\subsection{Automotive Mediated Reality Pipeline}
\label{ssec:pipeline}
\begin{figure*}[t]
    \centering
    \includegraphics[width=1\linewidth]{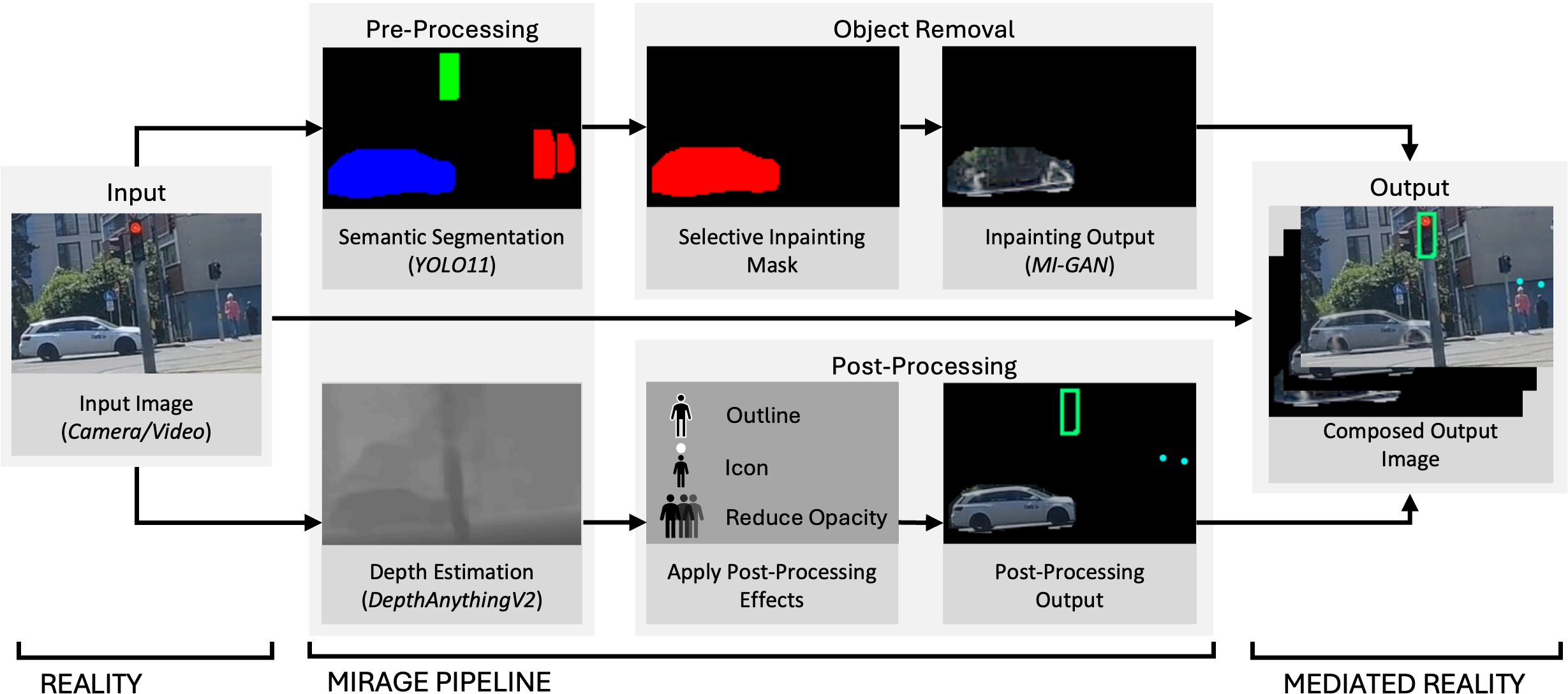}
    \caption{\tool processing pipeline. The input image is first pre-processed for semantic segmentation using \textit{YOLO11} and depth estimation using \textit{DepthAnythingV2}. Next, an object removal stage computes a selective inpainting mask, and the \textit{MI-GAN} model generates replacement pixels for targeted objects. Simultaneously, post-processing algorithms create the selected visualizations from the input and pre-processed data. Finally, all outputs are combined with the original image.}
    \label{fig:pipeline}
    \Description{A flow diagram labeled “MIRAGE Pipeline” shows how an input image from a camera or video is processed into a final “Mediated Reality” output. The top row illustrates “Pre-Processing” with semantic segmentation (YOLO11) to detect objects (shown in color masks), followed by “Object Removal” through a selective inpainting mask and the MI-GAN model. The bottom row shows depth estimation via DepthAnythingV2 and post-processing effects (e.g., Outline, Icon, Transparentize). Finally, both outputs merge into a composed output image that reflects AR, DR, or ModR effects applied to the original scene.}
\end{figure*}

Camera input is processed through \tool’s pipeline (see \autoref{fig:pipeline}), integrating object detection, segmentation, depth estimation, inpainting, and postprocessing via scripts and compute shaders to generate AMR visualizations. Each step runs as a coroutine in Unity to prevent blocking and lag. The pipeline can run sequentially, reducing GPU load but at a slower pace or in parallel, which speeds processing at the cost of higher GPU usage and allows different update rates per model. This setting is configurable in Unity.

\subsubsection{Computer Vision Models}
\label{sssec:models}
\tool uses \textit{Unity Inference Engine 2.2.1}~\cite{noauthorunitynodate} to run models in the ONNX\footnote{\url{https://onnx.ai/}, accessed: 06.02.2026} format directly within the Unity runtime. This eliminates the need for external interfaces to run models (e.g., a Python application).
Models were selected based on \href{https://docs.unity3d.com/Packages/com.unity.ai.inference@2.2/manual/supported-operators.html}{Inference Engine compatibility} and inference speed. \tool provides abstract runner scripts for easy model integration. Pre- and post-processing must be implemented manually.

First, the image is passed to a semantic segmentation model that classifies and generates masks for each object. These masks provide an \textit{object type} parameter used to apply AMR effects selectively. We use \textit{YOLO11s-seg}~\cite{yolo11_ultralytics}, trained on the \textit{COCO} dataset~\cite{lin_microsoft_2015} to include automotive-relevant object classes, though \tool also supports custom YOLO models (e.g., trained on \textit{Cityscapes}~\cite{Cordts2016Cityscapes}). The model, exported to ONNX using default Ultralytics settings, accepts a 640×640 input; thus, we downscale an input image while preserving its aspect ratio.

Additionally, we included the zero-shot absolute depth estimation model, \textit{DepthAnythingV2}~\cite{depth_anything_v2}, which generates a metric depth map. 
We used the \textit{depth anything v2 ViT-s outdoor dynamic} ONNX model\footnote{\url{https://github.com/fabio-sim/Depth-Anything-ONNX}, accessed: 06.02.2026}. Although it supports dynamic input sizes, we use the exact scaled-down dimensions as \textit{YOLO11-seg}. This alignment lets us combine outputs to compute each detected object's average or minimum distance. \tool uses distance as a second parameter for post-processing effects to display or adjust them based on range.

For DR, we implemented the \textit{MI-GAN}~\cite{sargsyan_migan_2023} inpainting model to remove objects selectively. It receives the input image and mask from \textit{YOLO11-seg}, removes the masked areas, and can be combined with a transparentizing effect. Additionally, it serves as a placeholder for future camera data. We exported the model to ONNX following the authors’ GitHub instructions\footnote{\url{https://github.com/Picsart-AI-Research/MI-GAN}, accessed: 06.02.2026} with minor modifications to omit pre- and post-processing, using a 512×512 input tensor. The updated export script is available in our GitHub repository.

The depth estimation and inpainting model are disabled by default and enabled as needed to improve performance. Our pipeline attempts to execute all model layers within one frame by default, ensuring fast frame updates. Finally, post-processed outputs are upscaled to the original input dimensions.

\begin{figure*}[t]
    \centering
    \includegraphics[width=1\linewidth]{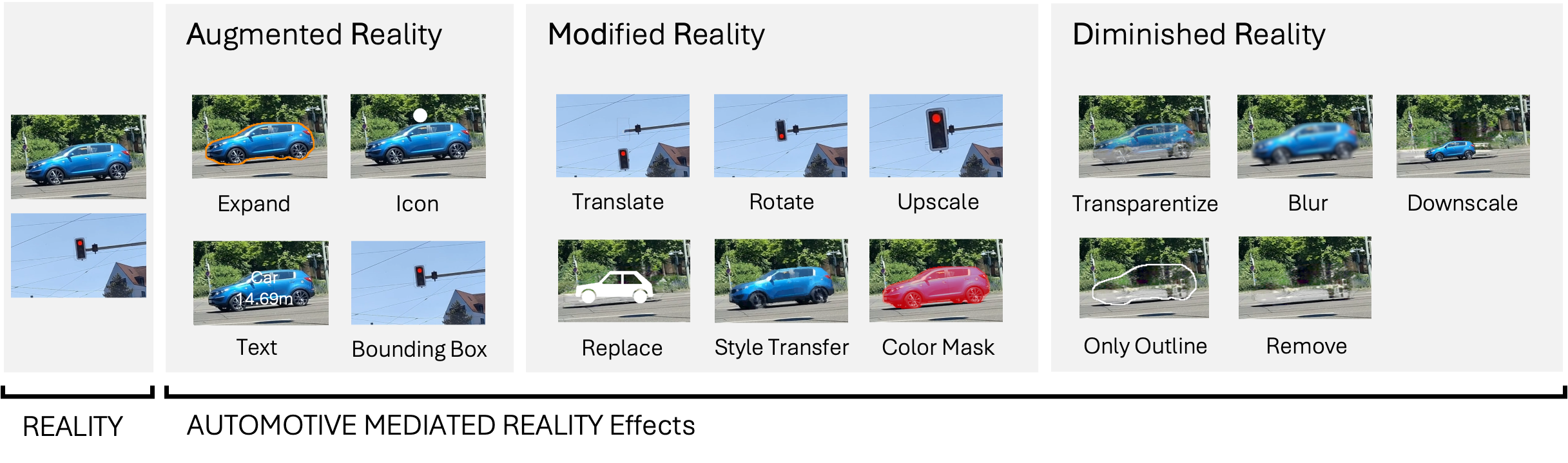}
    \caption{Overview of the 15 AMR effects we implemented for \tool applied to a car and traffic light in real-time. The effects \textit{Augment}, \textit{Diminish}, and \textit{Modify} the visual appearance of these objects.}
    \label{fig:effects}
    \Description{A horizontal illustration compares a real car and traffic light on the left with various AMR effects on the right. Under “Augment,” images show the car highlighted by an Outline, Icon, Text, or Bounding Box. “Modify” shows Translate, Scale, Rotate, Replace (changing the car to another object), Color Mask, and Style Transfer; “Diminish” features Transparentize, Downscaling, Only Outline, Blur, and Remove. This visually demonstrates how AMR can alter an object’s appearance or remove it entirely.}
\end{figure*}

\subsubsection{Application of Automotive Mediated Reality Effects}
\label{sssec:postprocessing}
Using a modular design that lets expert users adjust effects dynamically for each object type, we implemented 15 AMR effects (see \autoref{fig:effects} and the Video Figure) across different SAE levels. Nine effects were initially implemented (\textit{Bounding Box, Blur, Icon, Text, Color Mask, Transparentize, Outline, Style Transfer, and Remove}), and the remaining six were added after the user study based on the experts' feedback. Users can set visibility ranges and fine-tune effect-specific parameters to achieve the desired outcome.

\textbf{AR Effects} -  
We implemented four AR effects. The \textit{Outline} effect draws a colored border around each object’s segmentation mask. In contrast, the \textit{Bounding Box} uses YOLO11~\cite{yolo11_ultralytics} bounds to draw a rectangle that indicates object dimensions at a coarser granularity, which could be used to reduce cognitive load. The \textit{Icon} effect places a customizable icon at a specified position on the object, and \textit{Text} displays text with the object type (e.g., “person”) from YOLO11 and its distance from DepthAnything~\cite{depth_anything_v2}. \textit{Icon} and \textit{Text} contents can be replaced with any virtual content.
These AR effects target Dynamic/Location-Specific Elements > Highlight/Depict. They are useful for guiding driver attention (e.g., in SAE level 3 takeovers) and for communicating an AV’s road-user detection capabilities at higher automation levels (SAE 4–5). In these cases, AR can improve SA, trust, and perceived safety by directing attention to safety-critical cues (SA L1) or by externalizing the AV’s situational interpretation (SA L2), such as highlighting a cyclist emerging from occlusion or annotating a detected hazard. Prior work shows that outlining, bounding boxes, and icon overlays can improve detectability (SA L1) and help end-users interpret system reasoning (SA L2)~\cite{colley_effect_2020, colley_effects_2024}. According to our design space (see \autoref{tab:designspace}), AR is also useful for Action Elements > Depict/Combined to foster SA L3 (e.g., predicted road user trajectories \cite{colley_effects_2024} and pedestrian intention icons above their head \cite{colley_effect_2020}). AR, however, risks clutter and should be applied selectively to avoid competing with unmediated scene information (impeding SA and appraisals, such as trust \cite{colley2021effects}).

\begin{figure*}[t]
    \centering
    \includegraphics[width=\linewidth]{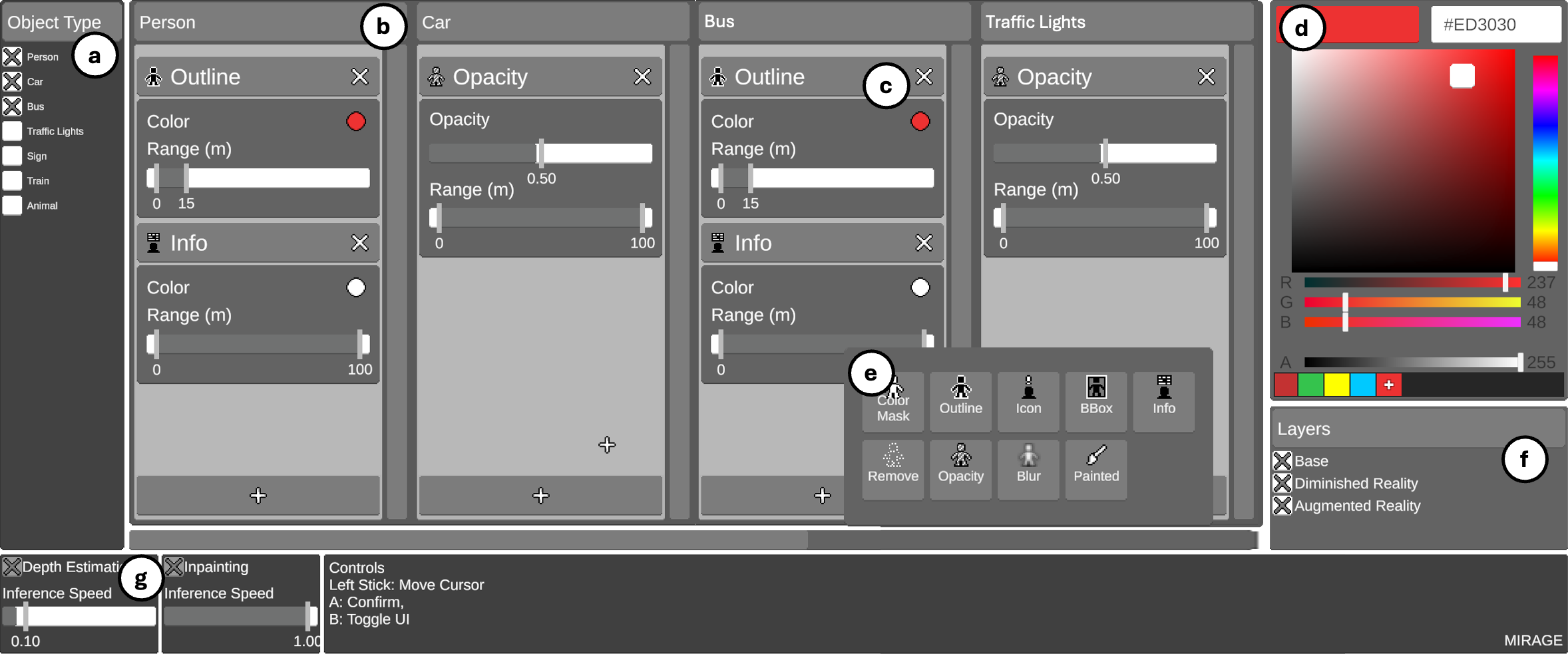}
    \caption{\tool effect control UI for researchers and practitioners (expert users). On the left (a), object types can be enabled or disabled. For each enabled type, a panel appears in the middle container (b), holding effect panels (c) with options adjustable via sliders and a color picker (d). Effects are added by clicking the “Add” (+) button to open a pop-up and select an effect (e). Layer visibility is toggled via a panel on the right (f), and model inference speed is adjusted at the bottom (g).}
    \label{fig:ui-showcase}
    \Description{A user interface is shown with four horizontal panels for different object types (e.g., Person, Car, Bus, Traffic Lights). Each panel (labeled “b” in the middle) includes effect controls such as Outline, Transparentize, and Info, adjustable by sliders and numeric fields. To the right (labeled “d”) is a color picker with hue and brightness bars, and below are layer visibility toggles (Base, Diminished Reality, Augmented Reality). Along the bottom, model settings for Depth Estimation, Inpainting, and Inference Speed are displayed.}
\end{figure*}

\textbf{DR Effects} - 
We implemented five DR effects. The \textit{Remove} effect deletes specific objects, potentially enhancing trust and safety in higher SAE levels~\cite{colley_feedback_2022}. \tool allows combining effects; for example, a “diminished outline”—as discussed by \citet{cheng_towards_2022} and originally explored by \citet{taylor_diminished_2020}—can be created by pairing the \textit{Outline} and reduction effects. The \textit{Transparentize} effect makes an object semi-transparent using inpainting~\cite{cheng_towards_2022}, and \textit{Downscale} overlays a scaled-down version at the original position. Lastly, the \textit{Blur} effect applies a Gaussian filter (with adjustable radius and sigma) to soften object details and guide attention~\cite{kosara_semantic_2004}. 
These DR effects target Dynamic/Location-Specific Elements > Highlight/Depict, while being less applicable to Action Elements, as they diminish existing information. They are useful for reducing irrelevant detail or background clutter—for example, softening building facades so pedestrians remain visually dominant, or reducing the salience of parked cars that distract from crossing traffic. Prior automotive work shows that stylistic abstraction~\cite{colley_effects_2022}, schematic renderings~\cite{wintersberger_explainable_2020}, and simplified scene variants~\cite{kraus_more_2020} can reduce cognitive demand. Yet DR carries the highest perceptual risk: diminishing or removing physically present actors may impair what users can perceive or comprehend (SA L1–L2), particularly when segmentation or inpainting errors create omissions, distortions, or spatial inconsistencies~\cite{calvi_effectiveness_2020, colley_feedback_2022}.

\textbf{ModR Effects} -
Our ModR effects modify existing elements without introducing new ones. Thus, they target Dynamic or Location-Specific Elements > Depict/Combined, while being less applicable to Action Elements. \textit{Color Mask} overlays a uniform color, similar to trust-enhancing semantic visualizations~\cite{colley_effects_2021}. \textit{Style Transfer} applies anisotropic Kuwahara filtering~\cite{kyprianidis_image_2009} with adjustable parameters~\cite{castillo_son_2017, kurzman_class-based_2019, lin_image_2021}, abstracting non-critical surfaces such as buildings. After object removal, objects can be \textit{Replaced} with symbolic graphics and then \textit{Scaled}, \textit{Translated}, or \textit{Rotated} to express relevance, predicted dynamics, or simplified structure—for example, enlarging a pedestrian approaching the curb, or rotating a vehicle icon to signal anticipated turning behavior. ModR is useful for calming overstimulating environments, creating schematic groupings, or making implicit semantics explicit. Prior DR work shows that suppressing irrelevant roadside elements supports focus~\cite{bauerfeind_navigating_2021}. However, ModR also alters spatial or motion cues: even minor deviations in shape, size, or alignment may disturb relational judgments and future-state projections (SA L2–L3).

\textbf{Adding Custom Effects} –
Additional AMR effects are feasible across all categories, such as reduced saliency, desaturation, and contrast~\cite{cheng_towards_2022}. \tool’s modular design and abstraction layers enable expert users to create custom effects. We provide abstracted C\# classes for compute shaders, CPU-based algorithms, and instructions in our open-source code repository.


\begin{table*}[ht!]
\centering 
\caption{Worst-case scenario benchmark results for MIRAGE with all effects enabled simultaneously. Mean ± SD by Hardware.}
\Description{A table depicting the worst-case-scenario benchmark results for MIRAGE. It shows the hardware used, the segmentation, inpainting, post-processing, and depth estimation time in ms, as well as the Unity FPS. The key takeaway is that with more modern, improved hardware, performance increases notably.}
\label{tab:hardware_means}
\centering
\resizebox{\ifdim\width>\linewidth\linewidth\else\width\fi}{!}{
\begin{tabular}[t]{lccccc}
\toprule
Hardware & Segmentation ms & Inpainting ms & Post Processing ms & Depth Estimation ms & Unity FPS\\
\midrule
\cellcolor{gray!10}{11th Gen Intel(R) Core(TM) i7-11700K @ 3.60GHz:NVIDIA GeForce RTX 3080} & \cellcolor{gray!10}{133.66 ± 33.45} & \cellcolor{gray!10}{146.78 ± 35.07} & \cellcolor{gray!10}{0.11 ± 0.12} & \cellcolor{gray!10}{148.48 ± 30.99} & \cellcolor{gray!10}{20.77 ± 10.95}\\
AMD Ryzen 5 9600X 6-Core Processor :NVIDIA GeForce RTX 4070 Ti SUPER & 75.52 ± 20.19 & 85.18 ± 19.78 & 0.05 ± 0.12 & 85.30 ± 19.64 & 28.48 ± 9.59\\
\cellcolor{gray!10}{Intel(R) Core(TM) i7-14700F:NVIDIA GeForce RTX 4080 SUPER} & \cellcolor{gray!10}{88.53 ± 25.38} & \cellcolor{gray!10}{105.43 ± 28.35} & \cellcolor{gray!10}{0.07 ± 0.10} & \cellcolor{gray!10}{106.90 ± 25.37} & \cellcolor{gray!10}{28.92 ± 14.28}\\
Intel(R) Core(TM) i7-14700KF:NVIDIA GeForce RTX 4090 & 72.39 ± 15.41 & 86.41 ± 20.33 & 0.04 ± 0.09 & 87.37 ± 17.93 & 28.43 ± 12.16\\
\cellcolor{gray!10}{Intel(R) Core(TM) i7-14700K:NVIDIA GeForce RTX 5080} & \cellcolor{gray!10}{57.50 ± 15.84} & \cellcolor{gray!10}{71.87 ± 23.75} & \cellcolor{gray!10}{0.04 ± 0.09} & \cellcolor{gray!10}{72.23 ± 23.20} & \cellcolor{gray!10}{38.89 ± 17.32}\\
\bottomrule
\end{tabular}}
\end{table*}

\subsection{Effect Control User Interface}
\label{ssec:ui}

To facilitate setup AMR effects, we built a UI (see \autoref{fig:ui-showcase}) using the \textit{Unity UI Extensions} package\footnote{\url{https://github.com/Unity-UI-Extensions/com.unity.uiextensions}, accessed: 06.02.2026}.
This UI is intended for expert users during prototyping as passengers; it is not designed for use during manual driving or by end-users in real-world driving scenarios.

A list on the left displays object types based on COCO~\cite{lin_microsoft_2015} classes from the segmentation model. During our expert user study, the object classes were pre-set based on their relevance to the automotive context and occurrence along the route. When a class is enabled, a vertical scroll panel appears; clicking the “add” (+) button opens a pop-up to add a post-processing effect panel, each containing settings adjustable via sliders or a color picker\footnote{\url{https://github.com/judah4/HSV-Color-Picker-Unity}, accessed: 06.02.2026} on the right. Visibility toggles for AR, DR, and Base layers let users quickly disable effects or make the windshield fully see-through to compare the environment with the video feed. In addition, AMR effects are directly applied to the real world instead of a live video feed. However, this option is more experimental as it makes the post-processing delay and the perspective offset between the virtual overlay and the real object more noticeable and may cause issues with specific effects such as \textit{Transparentize}. 

At the bottom of the UI, the currently active computational models (e.g., depth estimation, inpainting) are displayed; these change automatically based on the selected effects. Inference speed can also be adjusted: by default, \tool runs each model every frame, but models like depth estimation can be set to run every 10th frame to reduce GPU load. The UI supports mouse, keyboard, and gamepad input for control while seated in the vehicle.

\subsection{Technical Performance}
\label{ssec:performance}

We evaluated \tool using seven 1280x720@30 videos, taken from a diverse set of use cases, including driving in different cities, but also being in an office. Each video was run three times. They differed in length, from 3 min to 27:13 min (\m{8.95}, \sd{8.60}). We activated \textbf{all} effects and models to trial the most demanding setup. Benchmarks were run with Unity version \textit{6000.2.2f1} with a Quest 3. As all models are run in parallel, they compete for the same computational resources. If all models are executed each update cycle, the faster models (segmentation, inpainting) are limited to the performance of the slower, more demanding model (depth estimation). To optimize the pipeline's performance, we can control a model's inference speed/update rate. It controls the percentage of a model's layers that are processed in a single Unity update cycle. We set the inference speed of segmentation and inpainting to 100\%, meaning they are calculated in a single update cycle. As the depth estimation model is more demanding, we set its inference speed to 10\%, spreading its computation across 10 cycles. This reduces the number of outputs of the depth estimation per second while increasing the outputs for the other models.
We tested 4 computing hardware configurations with Quest Link (see \autoref{tab:hardware_means}). For analysis, we removed extreme outliers (4 $\times$ IQR). 
Overall, post-processing effects have a negligible performance impact. We can see an improvement with newer hardware, showing that \tool is usable and scales with hardware performance. 
It is important to note that the Unity FPS solely refers to the updated content; passthrough capabilities run without interference at the headset's set frequency (90 Hz for the Varjo 3).

While running the models in parallel improves inference speed, this optimization inherently introduces trade-offs in perceptual fidelity: The resulting frames may integrate depth or semantic estimations produced at different times, potentially compromising situational accuracy and raising safety concerns - particularly in scenarios where outdated inference outputs (e.g., depth estimation) are utilized for real-time decision-making.
Moreover, as with any machine learning-based system, incorrect outputs remain possible, including misclassification through YOLO11, depth estimation inaccuracies, and the hallucination of non-existent visual features during inpainting. Such errors may lead to incorrect or failed visualizations. While \citet{neumeier_TeleoperationHolyGrail_2019} identify 300 ms as a threshold at which driving performance begins to deteriorate—a latency \tool can remain below—other work shows that much lower limits matter for perceptual fidelity. \citet{mania_PerceptualSensitivityHead_2004} report that delays beyond 15–20 ms already degrade performance and immersion, and VR research recommends keeping interaction latency below 20 ms to avoid discomfort or motion sickness~\cite{hu_CellularConnectedWirelessVirtual_2020}.




\section{Expert User Study}
\label{ch:study}
To evaluate \tool and AMR, we invited experts from academia and industry to a study with the research question:
\begin{itemize}
    \itemrq \textit{How do automotive experts perceive \tool in terms of (1) usability, (2) task load, (3) and usefulness across use cases (manual driving, automated driving, other)?}
\end{itemize}

Therefore, participants rated both the anticipated utility of the \tool's outputs for drivers and passengers (end-users), as well as the utility as a prototyping tool for researchers and practitioners (expert users).

\subsection{Study Design}
\label{sec:study-design}
We collaborated with \textit{Mercedes-Benz Tech Motion (ITP/TM)}, who provided the test vehicle. The vehicle is equipped with a PC with a \textit{NVIDIA RTX 3090 Ti}, a \textit{Varjo XR3} HMD, an \textit{OptiTrack V120:Duo} tracking system, and an inertia measurement unit to allow for 6-DOF in-vehicle tracking. We additionally mounted a \textit{Logitech C920 HD Webcam} to the inside of the vehicle's windshield using a suction camera mount. Participants controlled the application with an \textit{Xbox One} controller. 
The study was conducted on and around the campus of Ulm University, Germany. 


\subsection{Measures}
\label{sec:measure}
We assessed the usability of \tool using the \textit{System Usability Scale} (SUS)~\cite{bangor_empirical_2008} and workload using  \textit{NASA Task Load Index} (NASA-TLX)~\cite{hart_development_1988}. We also asked participants four additional questions regarding \tool's use cases, which participants answered using 7-point Likert scales ranging from \textit{1 - Strongly Disagree} to \textit{7 - Strongly Agree}: \textit{(1) "This system would be useful for Manual Driving,"} \textit{(2) "This system would be useful for Automated Driving,"} \textit{(3) "This system would be useful to passengers in general,"} \textit{(4) "This system would be useful for other use cases (Pedestrian, at home, at work)."} Participants were asked to explain their opinions via a text box each time.
To gather feedback on each visualization concept, five items per concept were included, which participants answered using the same 7-point Likert scales: \textit{(1) "From a technical point of view, this visualization worked as expected,"} \textit{(2) "I would use this visualization concept in my professional work (projects/publications/etc.),"} \textit{(3) "I would use this visualization concept while driving manually."}, \textit{(4) "I would use this visualization concept as a passenger."}, \textit{(5) "I would use this visualization concept while driving in an automated vehicle."}
All settings participants changed while using \tool were logged.

Participants were also instructed to verbalize their thoughts continuously during the experiment, following the think-aloud protocol. Afterward, a short semi-structured interview was conducted with each participant, focused broadly on the experiment. Participants were asked about what they deemed positive and negative about \tool, as well as thoughts on \tool and AMR in general. We also asked about additional parameters that could be integrated into \tool in addition to \textit{object type} and \textit{distance}.

\subsection{Participants}
\label{sec:participants}
We invited N=9 experts with a background in automotive UI research from academia (n=5) and industry (n=4) to participate in our study. Participants' ages ranged from 24 to 37 (\m{27.78}, \sd{3.77}). Six participants identified as male, three as female. Eight participants held a master's degree, and one participant held a doctoral degree. Eight participants held a degree in the field of Information and Communication Technologies, and one participant held a degree in Social Sciences. All participants were currently employed in Germany. The experts' years of experience in the field of automotive UI ranged from 2 to 10 years (\m{4.11}, \sd{2.39}), and they were involved in 2 to 11 publications or projects (\m{7.00}, \sd{3.16}).

We intentionally recruited expert users rather than end-users (e.g., regular drivers) because \tool is targeted at designing and testing AMR concepts. Expert users can meaningfully judge technical feasibility, conceptual adequacy, perceptual risks, and integration issues. Their feedback, therefore, aligns with \tool’s intended stage of use: early-phase ideation, feasibility assessment, and identification of conceptual pitfalls before involving end-users in later controlled studies.

\subsection{Procedure}
\label{sec:procedure}
\begin{figure}[ht]
    \centering
    \includegraphics[width=.75\linewidth]{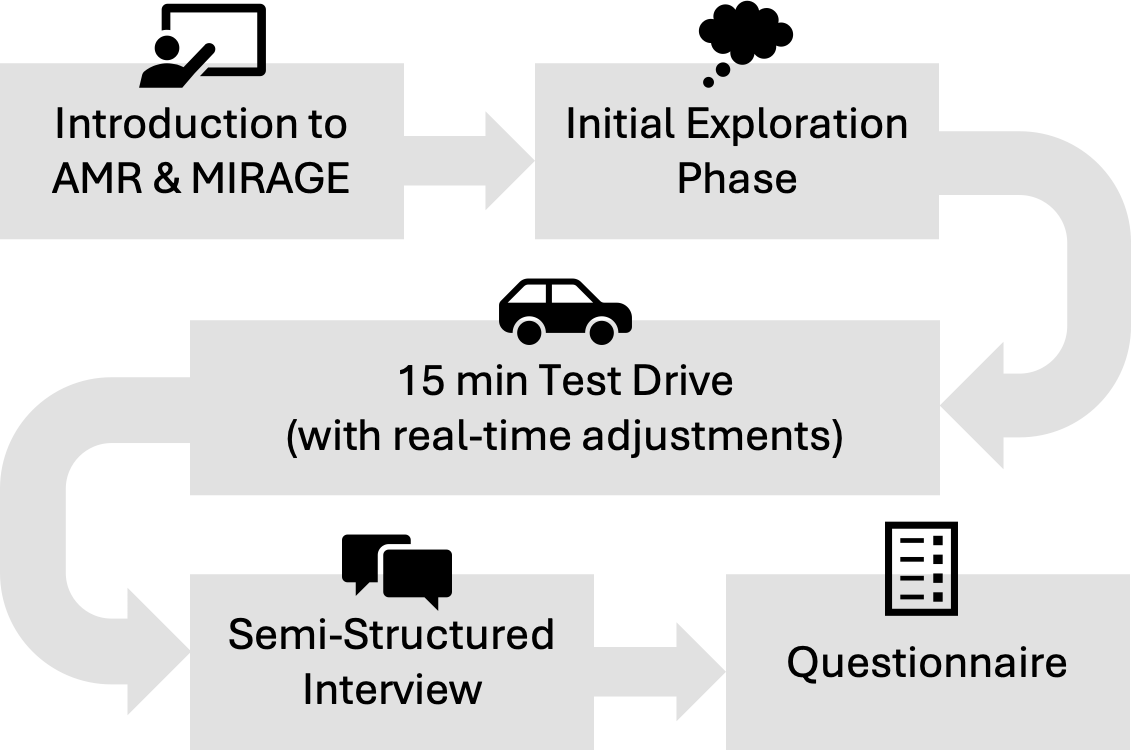}
    \caption{The procedure for the expert user study (N=9).}
    \label{fig:procedure}
    \Description{The figure shows the different steps for the expert users. Participants were first given an introduction to \tool and AMR, followed by an initial exploration phase where participants could freely explore \tool. Next followed a 15-minute test ride where participants could experience AMR first-hand. They were allowed to further adjust visualizations at any point during the test ride. The experience was followed up by a semi-structured interview and a questionnaire.}
\end{figure}

\autoref{fig:procedure} outlines the study procedure.
First, we provided a brief explanation of AMR and \tool's purpose: Enabling the exploration of AMR effects in real-world studies and scenarios (see Appendix \ref{appendix:instruction-texts}). Next, we introduced the participants to \tool by showcasing the nine available AMR effects so each was seen at least once. During the study, only nine effects were available to participants compared to the 15 effects shown in \autoref{fig:effects}. In the study, our focus was to cover the overarching AMR dimensions: AR, ModR, and DR. The remaining effects were created post-study based on the results and ongoing tool development.

Afterward, participants sat in the passenger's seat, wore the HMD, and received a brief explanation on navigating \tool's UI.
They then explored \tool freely without a time limit and could compose an initial set of AMR effects, while the vehicle was parked. Participants were instructed to communicate their thoughts continuously in a think-aloud approach. Once ready, the 15-minute test ride began, where they could experience the AMR effects they created first-hand. Participants did not perform driving tasks. The route near Ulm University was chosen to feature as many object types as possible. The following were present for all participants: cars, trucks, trams, traffic lights, street signs, and pedestrians. They were free to adjust their AMR effects at any time during the test ride (e.g., enabling/disabling AMR effects, or adjusting parameters). After the ride, participants were led inside the facility for a semi-structured interview about their thoughts on \tool and AMR as a whole. Finally, they answered a questionnaire about \tool and each AMR effect (accompanied by a video that showed the effect), followed by a demographics questionnaire. Participants were remunerated with €12.50. The study lasted around one hour. 

We used a free-exploration format rather than a structured task design because the goal of this early-stage evaluation was not to measure the performance effects of specific AMR concepts, but to understand how experts perceive the system’s capabilities, limitations, and design affordances when interacting with it in situ. This approach is commonly used in systems-focused HCI work to surface unexpected breakdowns, usage strategies, and conceptual issues that structured tasks may obscure \cite{Wobbrock.2016}.

\section{Results}
\label{ch:results}
In the following, we report the results of the expert user study. 

\subsection{Quantitative Data}
\label{sec:quantitative}

Quantitative ratings are descriptive and exploratory due to the small expert sample (N=9). They were not used to draw inferential conclusions but to complement the qualitative insights and contextualize experts’ subjective impressions of AMR effects.

\subsubsection{User Logs}
The logs created during the experiment gave us an overview of the effects that participants used for which \textit{Object Type} (see \autoref{tab:effects}). The \textit{Remove} effect was used by most participants (7, 0.77\%), followed by the \textit{Outline}, \textit{Text}, \textit{Transparentize}, and \textit{Blur} effects (5, 0.55\%). The \textit{Color Mask}, \textit{Icon}, and \textit{Bounding Box} effects were used four times (0.44\%), and the \textit{Style Transfer} effect was used the least (2, 0.22\%). The \textit{Person} and \textit{Car} object types were used by all nine participants, followed by \textit{Bus} (8, 0.88\%). The \textit{Animal} object type was only used by one participant (0.11\%).
Four of the nine participants (0.44\%) applied multiple effects to one object type (Participants P1, P2, P4, and P5). The effect range parameter was used by four (0.44\%) participants (P1, P2, P4, and P7).

\begin{table}[htbp]
  \centering
      \caption{AMR effects applied to different object types by participant ID. The last row and column show how many participants used an effect type or object type (denoted as $\Sigma$). Participants were shown all effects beforehand.}
      \Description{The table shows which AMR effect was applied to which object type by which participant. All nine participants applied effects to "Person" or "Car". Participants used all effects. "Remove" was used the most (7 times).}
  \resizebox{\columnwidth}{!}{
  \begin{tabular}{l|ccccccc|r}
    \hline
    \textbf{Effect} & \textbf{Person} & \textbf{Car} & \textbf{Bus} & \textbf{Traffic Lights} & \textbf{Sign} & \textbf{Train} & \textbf{Animal} & $\Sigma$ \\
    \hline
    Color Mask & P2, P4, P7 & --- & --- & P1, P7 & P1, P2 & --- & --- & 4 \\
    Outline & P0, P5 & P1 & --- & P4 & P3 & --- & P5 & 5\\
    Icon & P1, P2 & P2, P5 & P2, P5 & --- & --- & P0, P2 & --- & 4 \\
    Bbox & P8 & P8 & --- & P4, P5 & P2, P4 & --- & --- & 4\\
    Text & P3, P5, P8 & P2, P5, P7 & --- & --- & --- & --- & --- & 5 \\
    Remove & P6 & P0, P1, P3, P5, P6 & P4, P6, P7 & P6 & --- & P8 & --- & 7 \\
    Transparentize & --- & P4, P7, P8 & P1, P3 & --- & --- & P1, P7 & --- & 5 \\
    Style Transfer & --- & P0, P7 & P0 & --- & --- & --- & --- & 2 \\
    Blur & P1, P7 & P5, P6 & P3 & --- & --- & P3 & --- & 5 \\
    \hline
    $\Sigma$ & 9 & 9 & 8 & 5 & 4 & 6 & 1 &  \\
    \hline
  \end{tabular}
  }
  \label{tab:effects}
\end{table}


\subsubsection{System Feedback}\label{ssec:results-sus}
The resulting average score of \m{66.94} (\sd{13.79}) on the SUS indicated an average level of usability, according to \citet{bangor_empirical_2008}. The mean NASA-TLX~\cite{hart_development_1988} score was \m{7.77} (\sd{2.62}).
Participants rated whether they would see a use case for \tool in different scenarios on a 7-point Likert scale and explain use cases: Manual driving (\m4.56, \sd2.13), automated driving (\m5.56,\sd1.88), as a passenger (\m4.67, \sd1.87), and in other situations like as a pedestrian or at work (\m5.22, \sd1.72).

\subsubsection{Visualization Feedback Data}
\label{ssec:visfb}
\begin{figure}[ht]
\includegraphics[width=\linewidth]{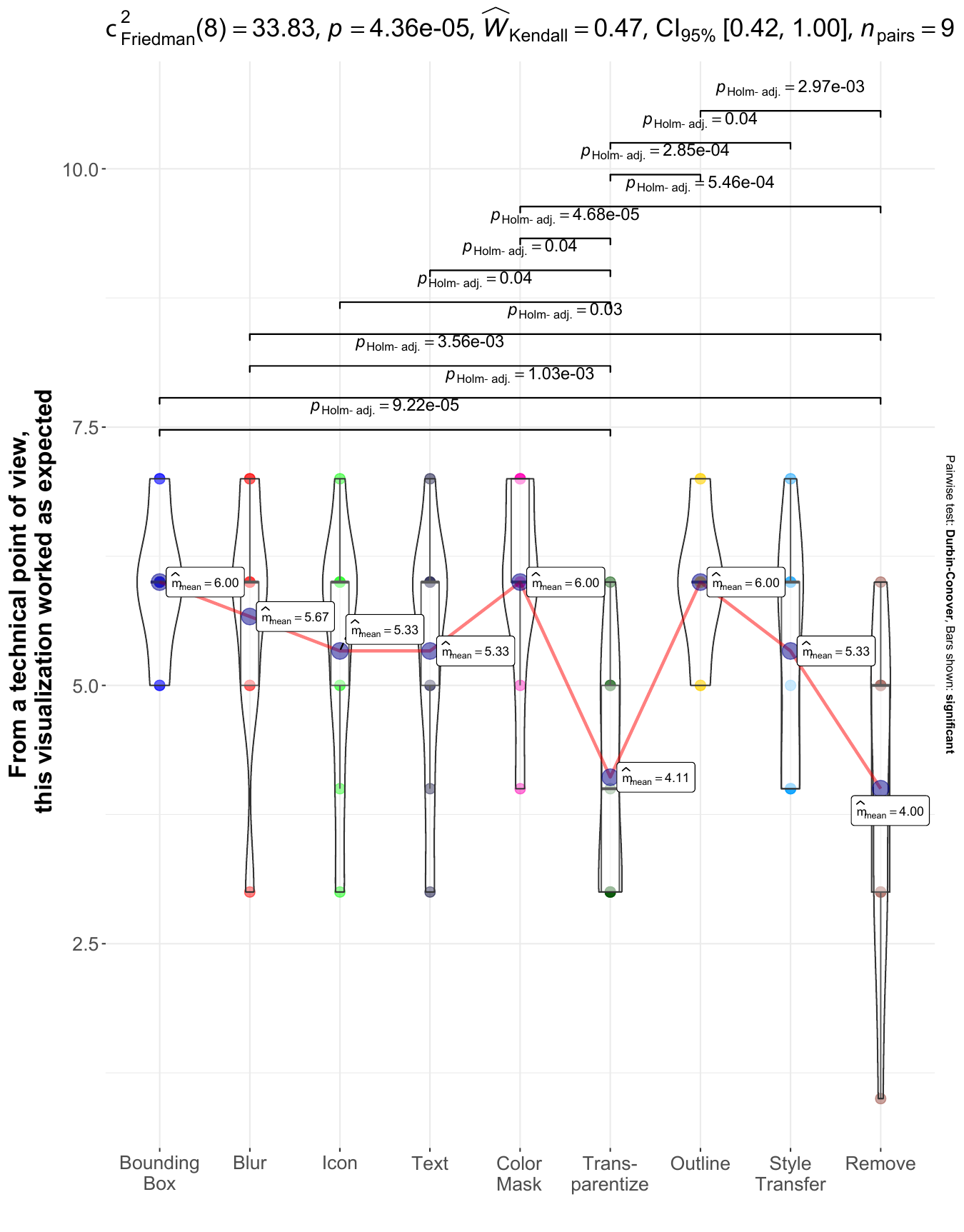}
\caption{Responses to the statement ''\textit{From a technical point of view,
this visualization} [AMR effect] \textit{worked as expected.}''}\label{fig:FB-technical}
\Description{A bar chart comparing various AMR effects shows mean ratings (mean) and Holm-adjusted p-values. The chart includes effects such as Bounding Box; Blur, Icon, Text, Color Mask; and Transparentize, Outline, Style Transfer, Remove, with mean values ranging from 4.00 to 6.00 and pHolm-adjusted values from 9.22×10⁻⁵ to 0.04. The x-axis is labeled with values 2.5, 5.0, 7.5, and 10.0. A pairwise Durbin-Conover test yields a Friedman statistic of 33.83, p = 4.36×10⁻⁵, and WKendall = 0.47 (95\% CI [0.41, 1.00]) based on 9 pairs, indicating significant differences as expected.}
\end{figure}

Participants were asked to rate statements for each visualization concept on a 7-point Likert scale. Regarding statement 1 \textit{From a technical point of view, this visualization worked as expected.}, the \textit{Bounding Box} (\sd{0.71}), \textit{Color Mask} (\sd{1.00}), and \textit{Outline} (\sd{0.71}) concepts were rated best at \m{6.00} (see \autoref{fig:FB-technical}). We found that all effects except the \textit{Remove} concept were rated significantly better than the \textit{Transparentize} concept. Further, the \textit{Bounding Box}, \textit{Blur}, \textit{Color Mask}, and \textit{Outline} concepts were rated significantly better than the \textit{Remove} concept.

For the statement \textit{I would use this visualization concept in my professional work (projects/publications/etc.)}, participants on average gave the highest rating to the \textit{Remove} concept (\m{5.22}, \sd{1.86}), followed by the \textit{Color Mask} concept (\m{5.00}, \sd{1.32}). The Icon concept was rated the lowest at \m{3.33} (\sd{2.00}). 
A Friedman test revealed no significant differences.


Participants gave low ratings for the statement \textit{I would use this visualization concept while driving manually.}, with \textit{Outline} (\sd{1.73}) and \textit{Bounding Box} (\sd{1.94}) receiving the highest ratings at \m{3.33} and the \textit{Blur} effect receiving the lowest (\m{1.67}, \sd{1.32}). 
A Friedman test revealed no significant differences.


For the fourth statement, \textit{I would use this visualization concept as a passenger.}, ratings were close together, but overall low with \textit{Remove} receiving the highest rating at \m{3.78} (\sd{2.39}) and the \textit{Icon} concept receiving the lowest (\m{2.56}, \sd{1.81}). A pairwise Student t-test revealed no significant differences. 


Lastly, for the statement \textit{I would use this visualization concept while driving in an automated vehicle.}, participants on average gave the lowest ratings to the \textit{Style Transfer} (\m{3.00}, \sd{2.06}) and \textit{Blur} (\m{3.22}, \sd{2.28}) concepts, while the \textit{Transparentize} (\m{4.67}, \sd{1.58}), \textit{Color Mask} (\m4.44, \sd{2.01}), \textit{Outline} (\m4.44, \sd{1.74}), and \textit{Remove} (\m{4.33}, \sd{1.80}) concepts received the highest ratings. A Friedman test revealed no significant differences. 


\subsection{Qualitative Data}
\label{sec:qualitative}
We recorded participants’ think-aloud feedback during the test ride, averaging \m{14:26 mins} (\sd{1:41 mins}), and conducted post-experiment interviews with all N=9 participants (average \m{8:53 mins}, \sd{1:34 mins}). The audio was transcribed using \textit{whisper}~\cite{radford_robust_2022} with manual corrections. We organized the feedback into key categories and reported anecdotal quotes summarizing the main points.

\subsubsection{Visualization Effects and Preferences}
\label{ssec:viseffepref}
The \textit{Remove} effect was discussed most (P0, P1, P3, P4, P5, P6, P8) with mixed reactions. While all participants deemed the concept interesting, many (P0, P1, P3, P6) pointed out that the technology is not yet ready, as the cognitive load may increase due to inpainting artifacts (P6). While P0 deemed the concept useful, the other participants (P1, P4) found no use cases. P6 mentioned they were already familiar with the concept but had previously perceived it completely differently when used in VR.
The \textit{Blur} effect was received positively by participants P1 and P7 during the interview, as it helped reduce distractions while maintaining awareness. Both participants further described the effect as "oddly satisfying". While participant P5 liked the \textit{Icon} effect, experts P0, P1, and P2 described it as useless, wishing for more customization options like changing the symbol (P0, P1) and changing the size (P2). The other AMR effects (\textit{Color Mask}, \textit{Outline}, \textit{Bounding Box}, \textit{Text}, \textit{Transparentize}, and \textit{Style Transfer}) received limited but positive feedback during the interviews. Having the ability to change the effect based on \textit{distance} was specifically highlighted by participants P5, P6, and P7.

\subsubsection{Technical Performance and Implementation}
\label{ssec:tech_perf_imp}
P1, P3, and P7 negatively highlighted low FPS and high system delay. P0 and P2 were positively surprised about the capabilities of the \textit{Yolo11} segmentation model. Further, detection inaccuracies were pointed out by P1, P2, P3, P6, and P7, stating that the system has to work well before being used (P3, P6). P0, P3, P5, and P6 positively highlighted the real-vehicle implementation, with three explicitly stating that on-road testing is necessary for evaluating AMR concepts. P6 emphasized that several effects are perceived “very differently” in real traffic than in VR, underscoring the limits of simulator-only evaluation. They further noted that \tool potentially improves prototyping of AMR concepts compared to traditional simulations (e.g., lab-based VR), as real-world deployment exposes practical issues, such as alignment errors, latency, and environmental variability.
No participant spontaneously reported visual discomfort or instability during use; however, the study did not include a dedicated assessment of cybersickness or perceptual side effects.

\subsubsection{UI and Interaction Design}
While the runtime customization capabilities of \tool were appreciated (P1, P6), multiple participants found the UI unintuitive (P1), which stopped them from trying out more things (P3, P4, P6). Participant P3 suggested using other input modalities, such as voice or gesture.

\subsubsection{Use Case Categories}

Participants listed potential use cases for \tool and AMR in general. We organized their use cases into five topics. P1, P2, P6, P7, and P8 mentioned \textbf{Safety Enhancement} in dangerous situations as one potential use case by highlighting critical information. 
Regarding DR, \textbf{Visibility Improvement} was mentioned by P0, P6, and P7, for example, on difficult routes (P0, P7) where the environment limits visibility (P7) or when overtaking (P6, P7). P7 also suggested removing the sun's glare to improve visibility. This is linked to the next topic: \textbf{Cognitive Load Management}. AMR could reduce distractions (P1, P5, P7) and mental workload (P8). \textbf{Context-specific Assistance} was also presented as a potential use case, e.g., when parking (P0, P7) near construction sites (P7) and railway crossings (P8).
Finally, \textbf{Experiential Enhancement} was identified: DR could be used for relaxation in AVs (P2, P7) or to reveal more of the environment by hiding people or traffic (P0, P1). Participant P0 also mentions using WSDs for entertainment by enabling games like "I spy".

Generally, P6 stated that a system covering the whole AMR spectrum and allowing customizability could be used for trust calibration in early AV adaptation. Further, it could be used for validation of many prior simulator-only studies.

\subsubsection{Use Case Feedback from Questionnaire}
Participants gave feedback for four use cases for \tool during the questionnaire.

They expressed mixed opinions about AR systems for manual driving. While some saw potential benefits in highlighting safety-critical information such as pedestrians, occluded objects, or dangerous situations, many expressed significant concerns. The reliability of such systems was a major issue - if not 100\% reliable, these augmentations could dangerously skew perception. Several noted that \tool could help manage attention by focusing drivers on relevant objects while reducing cognitive load and distractions. Specific use cases include visualizing parking spots, warning of crossing animals, and highlighting red traffic lights. However, multiple respondents emphasized that removing information or adding distracting visuals during manual driving could be dangerous, with one stating plainly that it would be "highly dangerous, neither having more information nor less would improve safety."

For automated driving, participants saw more potential benefits with fewer safety concerns. Key use cases included building trust by visualizing recognized objects to increase passenger trust in the system, providing explanations before requesting takeovers by showcasing the problems the system faces, supporting non-driving related tasks by removing surrounding ''noise'' like other traffic, and creating engaging infotainment systems when the car is handling driving tasks
One respondent noted the \tool "could aid in increasing passenger trust as it could visualize which objects were recognized properly, thus easing new end-users into the system." Others mentioned it could serve as a middle ground between manual driving and being a total passenger.

When using \tool for passengers, entertainment, and personalization were the primary benefits identified. Participants suggested the system could create personalized views of surroundings based on individual interests, remove crowds and vehicles that block the view to improve visibility, help passengers focus on non-driving tasks by removing distractions, or display contextual information such as points of interest.
One respondent said, "Passengers that are not interested could create 'their own' view of the surroundings." Several noted that safety features would be less necessary for passengers since they do not control the vehicle.

For use cases outside of the vehicle, users identified applications for pedestrian safety by highlighting approaching vehicles and clearing "blind spots." Home and work environments could be customized to reduce clutter and improve focus. Some expressed ethical concerns about social applications, particularly the ability to "remove" disliked people from one's perception, which could be "harmful to society."

\subsubsection{Contextual Adaptation Parameters}
We asked participants about additional parameters they would deem useful in addition to \textit{Object Type} and \textit{Distance}. Various \textbf{relevance parameters} have been suggested: Visualizing other vehicles should depend on their position on the road (P0, P6, P8), whether they are actively participating in traffic (P0, P8), as well as objects relation to the route (P0): Visualizing signs or traffic lights on other lanes may be irrelevant to the end-user. P5 and P7 also mention that highlighting traffic lights may only be relevant if they signal the driver to stop.
Further, three \textbf{environmental factors} were identified. AMR concepts could depend on the time of day (P1, P5, P7, P8): P7 and P8 suggest highlighting objects at night as they may be more difficult to see. Next, P2 and P5 mention the weather, as visibility may be affected, for example, by fog (P2). Participants P1, P2, P4, and P5 mention the vehicle's location. Pedestrian visualizations may be needed at bus or train stops (P8), and animal highlighting could be added when driving through a forest (P5, P7). 
The \textbf{Traffic Context} may also play a role, as visualizations could depend on traffic density (P4) and road type (P2, P4, P5). 
Next, participants identified \textbf{driver-specific factors}: P4 argues that eye gaze could only apply AR effects in the far peripheral field of view, assuming drivers are aware of objects they are directly looking at. P8 suggests driver fatigue could be a relevant parameter, as tired drivers may require more assistance. Further, they suggest a driver's experience level as a parameter to offer more assistance to inexperienced drivers.
Lastly, the \textbf{vehicle parameters} speed (P5, P7) and the driving mode/autonomy level (P4, P5, P6) were mentioned.

\subsubsection{Ethical Considerations}
Some experts voiced concerns about AMR and especially about DR. P6 described the experience of removing vehicles as scary, as suddenly, there is a disconnect between the end-user's and the vehicle's mental model of the situation. 
P5 mentions that DR can be used, but it should not fully remove objects from the environment. Regarding the \textit{Blur} concept, two P1 and P7 mentioned that it gives them a more anonymous feeling, as they were no longer aware if someone was looking at them.
Further, P7 raises questions about diminishing people against their will, creating a selective reality, a concept that could be abused.

\section{Discussion}
\label{ch:discussion}



With \tool, we provide a tool for expert users capable of covering the entire AMR spectrum (\textit{holisticity}). This includes AR (\textit{Outline, Bounding Box, Color Mask, Text, Icon}) and DR (\textit{Remove, Transparentize, Blur, Style Transfer}) effects, which have been used in prior works investigating their impact on perceived safety and trust~\cite{colley_feedback_2022}, SA and cognitive load~\cite{colley2021effects}, end-users reactions towards DR~\cite{cheng_towards_2022}, and object style transfer~\cite{kurzman_class-based_2019}. When participants rated the technical implementations, most effects received a positive rating, except for \textit{Transparentize} and \textit{Remove}, which received a neutral rating (see \autoref{fig:FB-technical}).

Moreover, particpants emphasized \tool’s value for AMR prototyping by enabling in-situ experimentation with AMR effects that are typically evaluated only in simulators. They reported that runtime parameter adjustments supported iterative exploration of design concepts, that real-vehicle use surfaced perceptual issues (e.g., alignment, latency, visual overload) that would remain hidden in VR, and that \tool could serve as an intermediate step between simulator studies and costly on-road deployments. While current technical constraints (see Section \ref{ssec:performance}) prevent \tool-created AMR effects from being used as an operational end-user system in traffic, the expert insights show that \tool may serve as a research tool that helps expert users identify conceptual, perceptual, and technical challenges early in the development process.

\begin{figure*}[t]
    \centering
    \includegraphics[width=\linewidth]{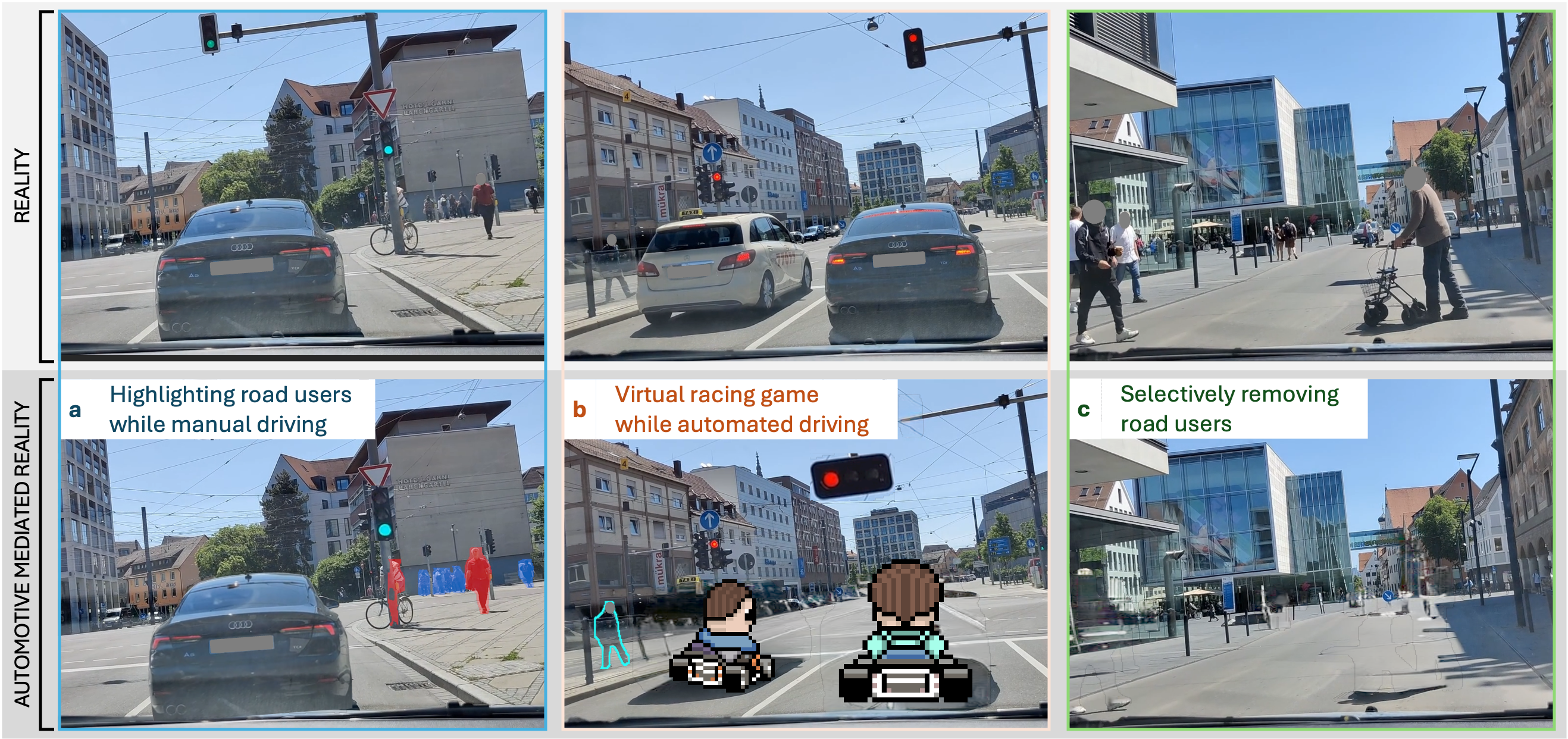}
    \caption{AMR use cases made with \tool. (a) To mitigate drivers overlooking vulnerable road users, pedestrians and cyclists can be highlighted with bright colors ((\textit{AR} > \textit{Expand}) × (\textit{Dynamic} > \textit{Highlight}), see \autoref{tab:designspace}), and traffic lights can be enlarged to enhance visibility ((\textit{ModR} > \textit{Upscale}) × (\textit{Location-Specific} > \textit{Highlight})). (b) In automated driving, non-driving tasks like entertainment are enabled: for example, vehicles can be replaced with carts ((\textit{ModR} > \textit{Replace}) × (\textit{Dynamic} > \textit{Depict})), and traffic lights can be modified—enlarged, repositioned, and rotated ((\textit{ModR} > \textit{Translate/Rotate/Upscale}) × (\textit{Location-Specific} > \textit{Highlight}))—to mimic racing lights for a gaming experience (see \citet{10.1145/3491101.3519741}). (c) However, ethical concerns arise if vulnerable road users, such as elderly pedestrians, are selectively diminished from reality ((\textit{DR} > \textit{Remove}) × (\textit{Dynamic} > \textit{Depict}).}
    \label{fig:use-case-examples}
    \Description{The image displays a 3×2 grid illustrating different use cases for Automotive Mediated Reality (AMR). The top panel shows a realistic driving scene in each column, while the bottom panel shows the same scene enhanced with AMR effects. In the first column, a busy urban street features vehicles, pedestrians, and cyclists; the enhanced view outlines the vulnerable road users and emphasizes their presence with bright, contrasting colors. In our design space, this would fall into the AR, Expand and Dynamic, Highlight cell. The second column shows a scene from an automated vehicle’s perspective; the enhanced view transforms conventional vehicles into stylized carts (Design space: ModR, Replace and Dynamic, Depict) and modifies traffic lights—enlarging, repositioning, and rotating them (Design Space: ModR, Translate/Rotate/Upscale and Location-Specific, Highlight) to evoke a dynamic racing interface for infotainment purposes. In the third column, a crowded intersection is depicted, and the enhanced version selectively diminishes certain elements, such as removing non-critical objects (Design Space: DR, Remove and Dynamic, Depict), which raises ethical concerns about the potential to obscure vulnerable road users. This grid effectively demonstrates how AMR can augment, modify, and diminish aspects of a driving environment to enhance safety, communication, and user experience.}
\end{figure*}

The suggestions regarding use cases for \tool were mostly in line with related work: Visibility improvement could be achieved through environmental cameras~\cite{lindemann_diminished_2017, lindemann_examining_2017, lindemann_acceptance_2019, rameau_real-time_2016, maruta_blind-spot_2021}. DR could reduce distractions~\cite{kim_dont_2020, katins_ad-blocked_2025} and cognitive load~\cite{colley_feedback_2022}, but has to work flawlessly. Other use cases include removing the sun's glare, as mentioned by \citet{hong_visual_2024, mann_mediated_1999}. \tool could support this by training \textit{YOLO}~\cite{Jocher_Ultralytics_YOLO_2023} to detect image glare and then applying post-processing effects that selectively reduce the brightness and contrast.
Context-specific assistance has also previously been explored: AR in automated parking scenarios leads to a higher user experience and SA~\cite{manger_explainability_2023}. For safety-critical scenarios, such as overtaking~\cite{manger_providing_2023}, additional information displayed through AR significantly increased driver trust and SA (e.g., see \autoref{fig:use-case-examples}a, constituting (\textit{AR} > \textit{Expand}) × (\textit{Dynamic} > \textit{Highlight}), see \autoref{tab:designspace}). AMR being used for experiential enhancement has been suggested by \citet{colley_feedback_2022}, where the surrounding environment was replaced to create a less stressful experience.
Moreover, in combination, AR, DR, and ModR concepts can create novel passenger AMR entertainment applications, such as a virtual racing game, during automated driving, as shown in \autoref{fig:use-case-examples}b (constituting (\textit{ModR} > \textit{Replace}) × (\textit{Dynamic} > \textit{Depict}), see \autoref{tab:designspace}) and \citet{10.1145/3491101.3519741}.

Moving beyond the automotive domain, \tool could be used in various use cases. For example,
\citet{katins_ad-blocked_2025} explored the idea of a real-life ad blocker that removes and replaces unwanted advertisements through MR. Custom-trained \textit{YOLO} models are supported by \tool. Thus, a model trained on billboard detection\footnote{e.g., \url{https://universe.roboflow.com/arslan-ongr8/billboard-xlvz1}, accessed: 06.02.2026} could be integrated quickly into the pipeline. 


In the following, we discuss \tool and the results of our expert user study. We give an outlook on ethical challenges, dark patterns, and potential AMR interaction from different viewpoints. Lastly, we talk about \tool's limitations and future research.

\subsection{Interaction with Automotive Mediated Reality}
Through enabling AMR, questions arise about how people will engage with it. 
Previous work~\cite{schramm_assessing_2023, sasalovici_bumpy_2025} evaluated in-vehicle interaction techniques for AR. Their findings show that gaze-based and head-based interactions led to low error rates as well as low perceived workload and high usability scores. We believe that these results partially translate to MR as a whole. However, a new question arises through the interaction of DR: 
\begin{quote}
\textit{How does an end-user interact with a diminished environment where objects were removed without their knowledge?}
\end{quote}

To ensure system transparency, we imagine a simple interaction (e.g., a voice command, a button, a gesture) that disables any MR experience at any point. 
However, end-users might become unaware that something has changed and might not naturally think about the simple interaction. Empirical work is needed here.

Furthermore, a new question regarding interaction with other road users arises. In traffic, drivers use certain non-verbal signals to communicate on the road. For example, drivers may flash their headlights to warn other road users of a hazard ahead. Similarly, people on the roadside may try to get a driver's attention in case of an accident, either for help or to warn them. End-users immersed in an AMR experience might miss these signals. Thus, the system should be able to interpret these signals, disable any DR visualization immediately, and potentially ask the end-user for further instructions. AR visualizations may need to be deployed as a counter-measure to increase end-users' SA quickly (see \autoref{fig:use-case-examples}a) while, potentially, immersed in entertainment applications (see \autoref{fig:use-case-examples}b). Future research must investigate DR's impact on SA in manual driving scenarios. This shows that various novel strategies are required to enable MR use cases.

So far, we have discussed AMR from the point of view of the person inside the vehicle. However, an interesting perspective is the role of other road users being augmented, manipulated, or diminished. Even if they can not see what is happening, should they have control or agency over AMR concepts applied to them? \citet{rixen_exploring_2021} found that even though visual alterations through AR are one-sided (only the wearer of an AR device can see the changes), people would like to be informed or even have "the last word" if an alteration is applied. Apple integrated outward-facing displays into their \href{https://www.apple.com/apple-vision-pro/}{Apple Vision Pro}, which show a virtual avatar of the wearer's face if the device's passthrough mode is enabled. If vehicles become AMR devices, there may be a need to communicate how the person inside the vehicle perceives the environment. This implies the possible need for external human-machine interfaces that communicate the vehicle's AMR state: 
\begin{quote}
\textit{Is passthrough enabled and the surroundings visible, are parts of the environment reduced, or is the environment entirely virtual?}
\end{quote}
Further, if there were an option to opt out of being augmented as a road user, it is unclear what that interaction would look like. 


\subsection{Dark Design Patterns for Automotive Mediated Reality}
With \tool, we enable the full MR spectrum inside vehicles (i.e., AMR). Despite limitations (see Section \ref{sec:limitations}), we must imagine what could be possible in the (near) future. While AMR could be used for safety enhancement, trust calibration, entertainment, and as a user experience, designers could manipulate reality in a way where the perceived world from inside the vehicle is always prettier than the real world, for example, by removing trash from the environment through inpainting. While this may sound harmless, the underlying implication is that a third party gains control over which parts of reality we perceive~\cite{colley_feedback_2022}. \citet{colley_feedback_2022} already partially discussed this topic from the perspective of automated vehicles in a section called "Driving in the Matrix: A Desirable Future?".

Ethical issues arise when the removed object is a person. While the idea of removing a crowd of tourists in front of a building to improve the sight-seeing experience may sound reasonable, the dystopian spin on this concept is the intentional removal of "unwanted" people from the perceived society (e.g., homeless people or political activists), creating a selective reality in the process (e.g., see \autoref{fig:use-case-examples}c, constituting (\textit{DR} > \textit{Remove}) × (\textit{Dynamic} > \textit{Depict}), see \autoref{tab:designspace}). \citet{kraus_what_2024} describe such behavior as \textit{Lying} or \textit{Disguising}, where they identified eight manipulative mechanisms/dark patterns for AR and VR: \tool could be used to nudge end-users towards taking certain routes by making other routes look worse than they are (\textit{Persuading}). AMR could strategically place virtual content along regular routes, making it difficult to avoid (\textit{Imbalancing Options}).

AMR could bait the end-user into using a different route, for example, for entertainment purposes, when in reality, the vehicle uses this detour for data collection (\textit{Baiting}, \textit{Directing Attention}, \textit{Requiring a Detour}). Lastly, if AMR is used for gaming, end-users could be pressured towards in-app purchases (\textit{Exploiting the drive to succeed}).

Privacy for other road users is another major concern regarding AMR use. Pedestrians need to be detected and segmented to be manipulated. Thus, people's data may (temporarily) be stored. In combination with metadata such as time of day, location, and weather, these vehicle systems could be abused for profiling.


Designers also need to be aware of the negative implications of AMR experiences. Aside from concerns regarding racism, more underlying negative stereotypes could be accidentally promoted through simple post-processing effects. A famous example is the "Mexican filter"~\cite{ponce_mexican_2021}, often used in media to tint scenes in Mexico with a yellow or sepia tone, which reinforces harmful stereotypes by making the country appear dirty, dangerous, or outdated. 


\subsection{Limitations \& Future Work}
\label{sec:limitations}

\tool has several limitations. First, \tool lacks object permanence and scene understanding, meaning objects are not persistently tracked or inferred between frames. This can result in inconsistencies when objects reappear or change position. Object tracking models could be integrated to solve this issue, for example, \textit{ByteTrack}~\cite{zhang_bytetrack_2022}.
Additionally, \tool does not feature 3D reconstruction, limiting its ability to represent depth and spatial relationships accurately. This task could be solved through computational models such as \textit{NeuralRecon}~\cite{sun_neuralrecon_2021}. Alternatively, a 3D point cloud could be created using the depth estimation model's per-pixel depth values and placing them in 3D space.

As a result, visual effects may not always align perfectly with the real-world scene. Although we aligned the perspectives as close as possible, using a camera that records from a different point always creates an offset. This becomes especially noticeable if the "true" passthrough mode is used, where the underlying video feed is disabled. We chose a separate camera feed to create a stable image from a fixed point that does not restrict HMD performance. Additionally, many manufacturers do not provide access to the HMD's camera feed without additional API licenses\footnote{\url{https://developer.apple.com/documentation/visionos/accessing-the-main-camera},\\ accessed: 06.02.2026}. Recently, Meta announced a \textit{Passthrough Camera API}\footnote{\url{https://developers.meta.com/horizon/documentation/unity/unity-pca-overview},\\ accessed: 06.02.2026} which allows for forward-facing camera access. However, this currently does not support the "Link" connection to a PC, which is necessary due to computational constraints.

Because the processing pipeline of \tool depends on real-time inference (segmentation, depth estimation, inpainting), temporary misclassifications, hallucinated content, or latency-related jitter may occur. Such artifacts can hide, distort, or spatially misalign traffic elements, making manual driving unsafe with the current prototype and potentially degrading the perceived stability or usefulness of AMR applications. In particular, generative inpainting may produce incorrect or incomplete background content when removing objects, which can momentarily obscure physically present actors (e.g., cyclists or pedestrians behind a car) or introduce distracting visual artifacts.

The depth estimation process currently creates the biggest bottleneck for \tool, as zero-shot depth estimation models are slow and compute-intensive, cutting \tool's performance in half without additional adjustments. Faster models in the future will most likely alleviate this issue. 

According to \citet{mori_survey_2017}, the core technical requirements for DR include background observation, scene tracking, region-of-interest detection, hidden view generation, and composition. All strategies are based on modifying or adding pixels to an image. While simpler visual transformations such as desaturation, reduced saliency, and reduced contrast could be implemented via compute shaders—abstract classes for which are already provided—more advanced methods like 3D object substitution~\cite{kari_transformr_2021} would require additional work due to \tool's lack of full 3D scene understanding. Similarly, "true" DR, which relies on multi-camera setups to reveal hidden views~\cite{lindemann_acceptance_2019, rameau_real-time_2016, maruta_blind-spot_2021}, is not currently supported.

\tool requires as little hardware as possible, making it usable in any vehicle (\textit{portability}). We partially achieved this goal as an RGB camera, an HMD, and a computer are needed. While we tested \tool with a \textit{Meta Quest 3} (2064x2208 pixels per-eye) and travel mode enabled, we used a more advanced setup for our expert user study with a \textit{Varjo XR3} (2880x2720 pixels per-eye) and \textit{Optitrack Duo} tracking system. However, compared to \textit{Portobello}~\cite{bu_portobello_2024} and \textit{XR-OOM}~\cite{goedicke_xr-oom_2022}, our setup is still less complex as it requires no additional sensors. However, depending on the use case, experts may need to integrate additional sensors into \tool: LiDAR could be used to replace inaccuracies caused by a depth estimation model. Facial or eye tracking could track driver fatigue, and an OBD-II connection could utilize vehicle data.

In our VR study, only the windshield overlay operated at a reduced frame rate, while the head-tracked base scene maintained a stable refresh rate of approximately 60 FPS. Low refresh rates increase cybersickness risk~\cite{Wang2023TVCG}, and because the overlay occupies a large portion of the field of view, even small mismatches between passthrough latency, head motion, and AMR overlays can produce discomfort or perceptual instability. Improving the overlay frame rate and reducing end-to-end latency remain necessary for longer usage durations and are essential before any safety-critical deployment.

The study reflects expert perspectives rather than those of regular drivers or passengers. This aligns with \tool’s purpose as a prototyping tool but limits claims about end-user experience. Future work should investigate holistic AMR concepts (i.e., AR, ModR, and DR) with end-users once the rendering stability of ModR and DR is improved. Prior work has already enabled manual driving with passthrough VR HMDs in real vehicles \cite{bu_portobello_2024, goedicke_xr-oom_2022}.

In the future, we will improve \tool and integrate the suggestions (e.g., contextual adaptation parameters) from the expert user study. To further evaluate the current state of MR, we will use \tool to recreate and conduct simulator-only studies.
Further, MR could reduce clutter and improve focus at home or work~\cite{cheng_towards_2022, yokoro_decluttar_2023}.



\section{Conclusion}
This work presents \tool, the first holistic open-source tool to deploy and evaluate MR concepts in a real vehicle. We simplify deploying and testing such ''AMR'' concepts on the road by simulating a WSD using an HMD and a camera in combination with a pipeline utilizing state-of-the-art computational models for object identification and segmentation, depth estimation, and inpainting. We implemented 15 AMR effects that can be applied to objects based on type and distance. \tool supports combining AMR effects (i.e., AR, DR, and ModR) to create more advanced AMR concepts.

We evaluated \tool in an expert user study (N=9). Participants appreciated its fidelity and flexibility, recognizing its value for trust calibration in AVs and validation of prior research. Despite their technical limitations, the implemented AR and DR concepts were well received, with a particular interest in DR. 
Key challenges identified include performance constraints, UI usability, and the need for flawless effects produced by computational models for distraction reduction. The modularity of \tool allows for customization and expansion, allowing expert users to adapt effects based on specific needs. 
We discussed opportunities for AMR use and beyond in the context of MR, including potential ethical challenges and hurdles. 
With \tool, we lay the critical foundation for real-world, real-time AMR experiences and research. 

\section*{Open Science}
The source code of \tool is available on our \href{https://github.com/J-Britten/MIRAGE}{GitHub repository}.
Additionally, this repository contains the technical performance benchmark data and evaluation script, as well as anonymized data from the expert user study.

\begin{acks}
We thank all study participants. 

\end{acks}

\bibliographystyle{ACM-Reference-Format}
\bibliography{chi26-references}

\appendix
\section{Expert User Study Introduction Text} \label{appendix:instruction-texts}
Translated from German:

\begin{quote}
\textit{Initial question: Are you familiar with the term mediated reality?}
\end{quote}

\begin{quote}
\textit{Depending on answer: Let participant describe their definition, then proceed:}
\end{quote}

\begin{quote}
\textit{You may be familiar with the reality-virtuality continuum by Milgram. On the one end, you have the real environment, on the other, the virtual environment, and in between, everything else, from augmented reality to augmented virtuality. It is essentially a spectrum that allows us to describe all possible combinations of real and virtual objects. However, if we look at the term "augmented reality", the word augment means to add or to expand something. But what if we want to modify an object by changing its color or by reducing it to transparency? AR does not cover that. And that is where mediated reality comes in. It is an umbrella term that adds a second dimension to the virtuality continuum, which allows us to describe what is done to an object. It covers the term AR and also includes modified and diminished reality. }
\end{quote}

\begin{quote}
\textit{Now, commonly, in automotive HCI research, many new mediated reality concepts are explored, but they are rarely evaluated in real vehicles. And that is where our tool MIRAGE comes in. It allows us to create and evaluate MR visualizations in a real vehicle and in real time. And to gather feedback on MIRAGE and the concept of in-vehicle MR, we invited you.}
\end{quote}

\begin{quote}
\textit{In this experiment, you can choose between nine effects and freely apply them to any object. And to get an idea of how they work, here is a short video that showcases each effect:}
\end{quote}

\begin{quote}
(Showing video clip of each effect)    
\end{quote}


\end{document}